\documentclass[12pt,preprint2,appendixfloats]{emulateapj}
\usepackage{natbib}
\usepackage{amsmath}
\usepackage{graphicx}
\usepackage{wasysym}
\usepackage{txfonts}
\usepackage{xspace}
\usepackage{url}
\usepackage{textcomp}
\usepackage{multirow}
\usepackage{lipsum}

\newcommand{\gx}{GX~339$-$4\xspace}

\newcommand{\swift}{\textsl{Swift}\xspace}

\newcommand{\xmm}{\textsl{XMM-Newton}\xspace}

\newcommand{\suz}{\textsl{Suzaku}\xspace}

\newcommand{\xte}{\textsl{RXTE}\xspace}

\newcommand{\nustar}{\textsl{NuSTAR}\xspace}

\newcommand{\snr}{S/N\xspace}

\newcommand{\msun}{\ensuremath{\text{M}_{\odot}}\xspace}

\newcommand{\redchi}{\ensuremath{\chi^{2}_\text{red}}\xspace}

\newcommand{\feka}{\ensuremath{\mathrm{Fe}~\mathrm{K}\alpha}\xspace}

\newcommand{\nh}{\ensuremath{{N}_\mathrm{H}}\xspace}

\newcommand{\ledd}{\ensuremath{{L}_{\mathrm{Edd}}}\xspace}
\renewcommand{\deg}{\ensuremath{^\circ}}

\shorttitle{Complex accretion geometry of \gx }
\shortauthors{F\"urst et al.}

\begin{document}

\title{The complex accretion geometry of \gx as seen by \nustar and \swift}

\author{F.~F\"urst\altaffilmark{1}}
\author{M.~A.~Nowak\altaffilmark{2}}
\author{J.~A.~Tomsick\altaffilmark{3}}
\author{J.~M.~Miller\altaffilmark{4}}
\author{S.~Corbel\altaffilmark{5,6}}
\author{M.~Bachetti\altaffilmark{7,8,9}}
\author{S.~E.~Boggs\altaffilmark{3}}
\author{F.~E.~Christensen\altaffilmark{10}}
\author{W.~W.~Craig\altaffilmark{3,11}}
\author{A.~C.~Fabian\altaffilmark{12}}
\author{P.~Gandhi\altaffilmark{13,14}}
\author{V.~Grinberg\altaffilmark{2}}
\author{C.~J.~Hailey\altaffilmark{15}}
\author{F.~A.~Harrison\altaffilmark{1}}
\author{E.~Kara\altaffilmark{12}}
\author{J.~A.~Kennea\altaffilmark{16}}
\author{K.~K.~Madsen\altaffilmark{1}}
\author{K.~Pottschmidt\altaffilmark{17,18}}
\author{D.~Stern\altaffilmark{19}}
\author{D.~J.~Walton\altaffilmark{19,1}}
\author{J.~Wilms\altaffilmark{20}}
\author{W.~W.~Zhang\altaffilmark{17}}

\altaffiltext{1}{Cahill Center for Astronomy and Astrophysics, California Institute of Technology, Pasadena, CA 91125, USA}
\altaffiltext{2}{Massachusetts Institute of Technology, Kavli Institute for Astrophysics, Cambridge, MA 02139, USA}
\altaffiltext{3}{Space Sciences Laboratory, University of California, Berkeley, CA 94720, USA}
\altaffiltext{4}{Department of Astronomy, The University of Michigan, Ann Arbor, MI 48109, USA}
\altaffiltext{5}{Laboratoire AIM (CEA/IRFU - CNRS/INSU - Universit\'e Paris Diderot), CEA DSM/IRFU/SAp, 91191 Gif-sur-Yvette, France}
\altaffiltext{6}{Station de Radioastronomie de Nan\c{c}ay, Observatoire de Paris, CNRS/INSU, USR 704 - Univ. Orl\'eans, OSUC, 18330  Nan\c{c}ay, France}
\altaffiltext{7}{Universit\'e de Toulouse; UPS-OMP; IRAP; Toulouse, France}
\altaffiltext{8}{CNRS; Institut de Recherche en Astrophysique et Plan\'etologie, 31028 Toulouse cedex 4, France}
\altaffiltext{9}{INAF/Osservatorio Astronomico di Cagliari,  09047 Selargius (CA), Italy} 
\altaffiltext{10}{DTU Space, National Space Institute, Technical University of Denmark, 2800 Lyngby, Denmark} 
\altaffiltext{11}{Lawrence Livermore National Laboratory, Livermore, CA 94550, USA}
\altaffiltext{12}{Institute of Astronomy,  Cambridge CB3 0HA, UK} 
\altaffiltext{13}{Department of Physics, Durham University, Durham DH1 3LE, UK} 
\altaffiltext{14}{School of Physics \& Astronomy, University of Southampton, Highfield, Southampton SO17 1BJ, UK}
\altaffiltext{15}{Columbia Astrophysics Laboratory, Columbia University, New York, NY 10027, USA}
\altaffiltext{16}{Department of Astronomy \& Astrophysics, The Pennsylvania State University, University Park, PA 16802, USA}
\altaffiltext{17}{CRESST, UMBC, and NASA GSFC, Code 661, Greenbelt, MD 20771,USA}
\altaffiltext{18}{NASA Goddard Space Flight Center, Greenbelt, MD 20771, USA}
\altaffiltext{19}{Jet Propulsion Laboratory, California Institute of Technology, Pasadena, CA 91109, USA}
\altaffiltext{20}{Dr. Karl-Remeis-Sternwarte and ECAP,  University of Erlangen-Nuremberg, 96049 Bamberg, Germany} 

\begin{abstract}
We present spectral analysis of five \nustar and \swift observations  of \gx taken during a failed outburst in summer 2013. These observations cover Eddington luminosity fractions in the range $\approx$0.9--6\%. Throughout this outburst, \gx stayed in the hard state, and all five observations show similar X-ray spectra with a hard power-law with a photon index near 1.6 and significant contribution from reflection. Using simple reflection models we find unrealistically high iron abundances. Allowing for different photon indices for the continuum incident on the reflector relative to the underlying observed continuum  results in a statistically  better fit and reduced iron abundances. With a photon index around 1.3, the input power-law on the reflector is significantly harder than that which is directly observed. We study the influence of different emissivity profiles and geometries and consistently find an improvement when using separate photon indices. 
The inferred inner accretion disk radius is strongly model dependent, but we do not find evidence for a truncation radius larger than $100\,r_g$ in any model.  The data do not allow independent spin constraints but the results are consistent with the literature (i.e., $a>0$). 
Our best-fit models indicate an inclination angle in the range 40--60$^{\circ}$, consistent with  limits on the orbital inclination but higher than reported in the literature using standard reflection models.
The iron line around 6.4\,keV is clearly broadened, and we detect a superimposed narrow core  as well. This core originates from a fluorescence region outside the influence of the strong gravity of the black hole and we discuss possible geometries.
\end{abstract}


\section{Introduction}
Black holes accreting via Roche-lobe overflow from a low-mass companion star typically demonstrate a strongly transient behavior: while they spend most of their time in quiescence, during outbursts they can reach luminosities up to $\sim$$10^{39}$\,erg\,s$^{-1}$, the Eddington limit for a 10\,\msun black hole \citep[and references therein]{remillard06a}. The X-ray spectra can primarily be described by two components: a multi-temperature accretion disk and a hot electron gas corona. The corona  shows up as a power-law component, originating from Compton up-scattering of soft disk seed photons by coronal hot electrons.
Outbursts typically start in a so-called low/hard state in which a power-law with a photon index $\Gamma \leq 1.7$ dominates the X-ray spectrum.  The accretion disk contributes only weakly to the X-ray spectrum, with typical disk temperatures $\text{k}T\lessapprox0.2$\,keV. The disk is more evident as a reflector of the hard X-ray continuum, resulting in a characteristic iron K$\alpha$ line at 6.4\,keV and a Compton hump between 20--40\,keV \citep{ross05a}. In the later stages of the outburst, the source typically enters the high/soft state, in which the accretion disk gets hotter and becomes the dominant contributor to the X-ray spectrum, while at the same time the power-law becomes softer.

In the high/soft state strong evidence exists that the accretion disk extends all the way to the innermost stable circular orbit \citep[ISCO, ][]{shakura73a}. The radius of the ISCO depends strongly on the black hole spin and with it the distortions of the fluorescent iron line due to orbital motions and general relativistic effects. By modeling these distortions, it is possible to estimate the spin of the black hole.
An alternative method to measure the spin is to use the continuum flux from the disk black body component, which is also a function of the ISCO \citep{zhang97a,davis05a,mcclintock14a}.

In the low/hard state it has been postulated that the accretion disk is truncated and that the inner regions of the accretion flow are described by an optically thin, advection dominated accretion flow \citep[ADAF, see, e.g.,][]{narayan95a, esin97a, taam08a}.
The truncation radius is predicted to be at several $100\,r_g$, and therefore the fluorescent \feka line from disk reflection should be narrow, and the blackbody radiation from the disk should be relatively cold.

Measurements of the inner accretion disk radius at luminosities $>1\%\,\ledd$ in the low/hard state typically are consistent with, or even require, an accretion disk extending all the way to the ISCO \citep[see, e.g.][]{nowak02a, miller06a, petrucci14a, miller14a}. 
\citet{reis10a} extend these measurements down to 0.05\% $\,\ledd$  using a sample of 8 sources and find that there is still evidence for an accretion disk close to the ISCO. 
However, \citet{tomsick09a} measure a truncated  disk  with an inner radius $R_\text{in} >175\,r_g$ for \gx at $L_\text{x}\approx0.14\%\,\ledd$, 
assuming an inclination of $i=30^\circ$. This is the lowest luminosity of \gx for which such a measurement is available and is the first time a clearly truncated disk is seen.
 The authors speculate that the accretion disk moves quickly outwards at luminosities $<1\%\,\ledd$, but it is not clear what mechanism triggers this change in the accretion disk geometry. This idea of a recession below 1\%\,\ledd in \gx is supported by \citet{allured13a} who measure inner radii around $10\,r_g$ for luminosities between 0.5--5\%\,\ledd.
 
\gx is the archetypical transient black hole binary. It has been studied intensively since its discovery in 1973 \citep{markert73a} and shows a high level of activity, with an average of one outburst of varying strength every 2 years. 
During the hard-states it reaches fluxes in excess of 150\,mCrab in the \swift/BAT 15--50\,keV energy band. This high flux and semi-regular outburst activity makes \gx an ideal target to study the low/hard state in detail.

The optical companion is not directly observable, and the only measurement of the orbital period has been obtained by measuring Doppler shifts of fluorescent lines caused by the irradiation of the companion \citep{hynes03a}. These measurements give an orbital period $P=1.7$\,d and together with the upper limit on the companion's magnitude and assumed spectral type, \citet{zdziarski04a} estimate the black hole mass to be $M=10\,\msun$ and the distance $d=8$\,kpc. These values are consistent with the mass limit of $M\geq7\,\msun$ calculated by \citet{munozdarias08a} using a similar method and we adopt them throughout this paper.

The inclination of the binary orbit is only weakly constrained.  As \gx is a non-eclipsing system, the inclination has to be lower than $60\deg$ \citep{cowley02a}, and \citet{zdziarski04a} estimate from the secondary mass function a lower limit of $45\deg$. \citet{shidatsu11a} summarize all constraints and derive a best estimate of $\approx50^\circ$.

The spin of the black hole in \gx is currently under debate in the literature. Estimates from reflection modeling and disk continuum models do not yet give consistent results regarding  spin and inclination.  This discrepancy is likely due to different underlying assumptions in the models 
\citep[for recent reviews on both methods, see][and \citealt{mcclintock14a}]{reynolds14a}. However, both methods rule out a Schwarzschild (i.e. non-spinning) black hole with high significance \citep{kolehmainen10a, miller08a}.

\citet{miller04a, miller08a} use \suz  and \xmm data to measure the relativistic broadening of the \feka line outside of the low/hard state and obtain $a=0.93\pm0.05$. \citet{reis08a}, using \xmm and \xte data of both a soft and a hard state, confirm that value but require an inclination as low as $i\approx20^\circ$ of the inner accretion disk.  At these very low inclinations fitting the disk continuum to measure the spin gives consistent results, however, \citet{kolehmainen10a} argue that this requires a strong misalignment between the orbit's rotation axis and the spin axis, as the orbital inclination is limited to be $i_\text{orb} > 45^\circ$. \citet{kolehmainen10a} therefore prefer a disk inclination $>45^\circ$ which results in significantly lower spin ($a<0.9$).

\citet{plant14b} present \xmm data taken during the same  failed outburst in 2013 presented here (see below), and fit an inclination $33^\circ\pm3^\circ$ while fixing the spin at $a$=0.9. They find consistent results between the disk continuum method and the reflection modeling assuming this inclination, and measure a small but significant truncation of the disk, with an inner radius of 20--30\,$r_g$. They do not, however, address the misalignment between orbital and disk inclination.

Therefore both the inner disk inclination and the spin value remain under discussion and need further investigation with sensitive X-ray instruments, especially covering the hard X-rays to get a good measure of the continuum parameters.

As shown by \citet[and references therein]{corbel03a}, the radio flux of \gx is strongly correlated with the X-ray flux in the low/hard state, following the non-linear relationship $L_\text{radio} \propto L_\text{x}^{\sim0.7}$. This correlation has been connected to synchrotron radiation from the the jet also influencing the hard X-ray spectrum \citep{markoff03a}. \citet{corbel13a} have shown that the onset of an outburst is also observed in the radio spectrum, which switches from a negative spectral index $\alpha$, where the radio spectrum is described by $\nu^\alpha$, to $\alpha \approx 0.5$ during the low/hard state. This correlation therefore provides information about the connection between accretion and ejection and between the jet and the corona.

\begin{figure}
\includegraphics[width=0.96\columnwidth]{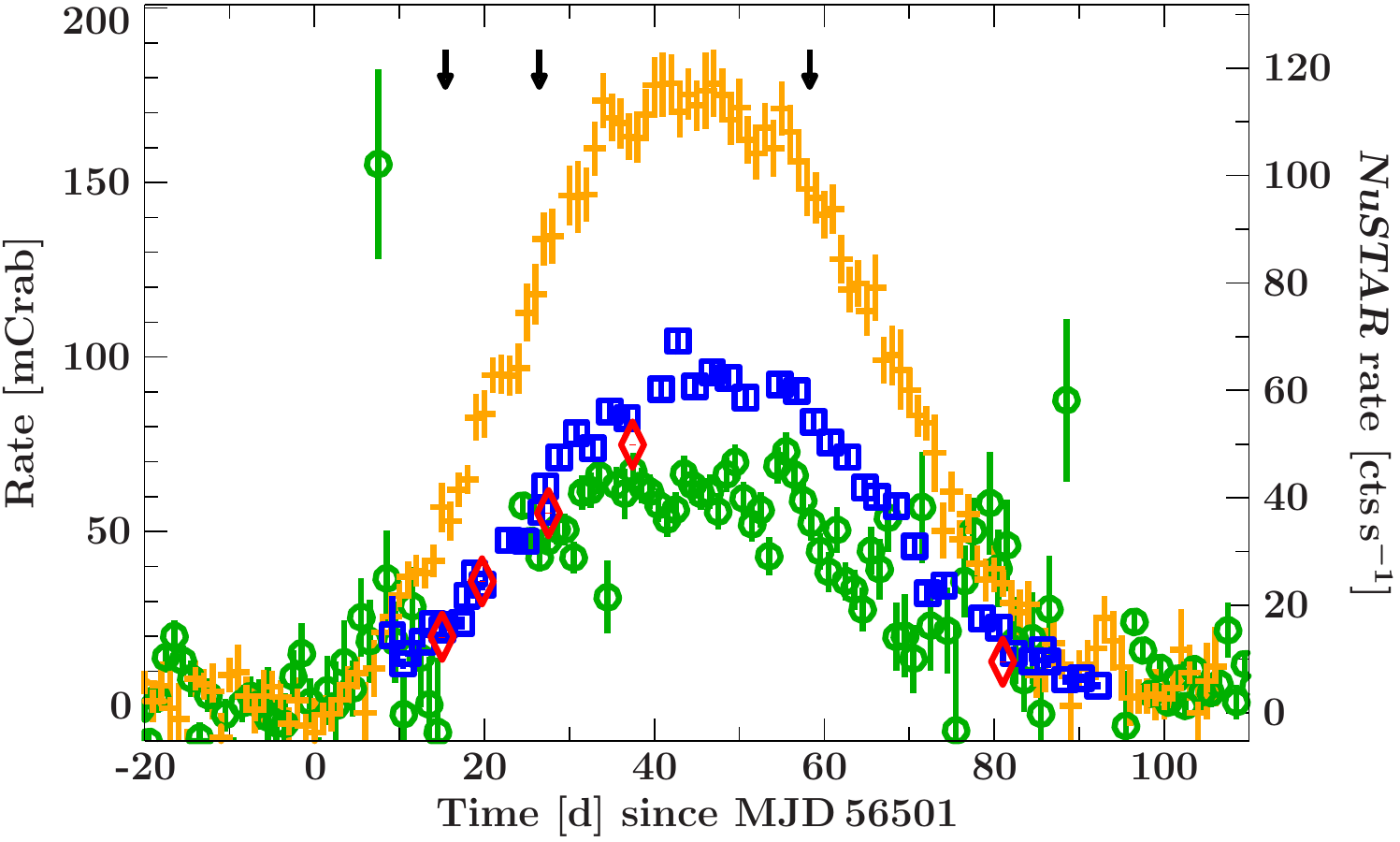} 
\caption{Light curves of \swift/BAT (15--50\,keV, orange crosses), MAXI (2--20\,keV, green circles), \nustar (3--79\,keV, red diamonds), and \swift/XRT (2--10\,keV, blue squares) of the 2013 outburst. All count-rates have been rescaled to mCrab fluxes in the respective energy band of the instrument. The right-hand $y$-axis gives the average  measured \nustar count-rates of each observation. The downwards-pointing arrows at the top indicate the times of the ATCA observations.}
\label{fig:batlc}
\end{figure}

In August 2013, optical and X-ray monitoring detected the onset of a new outburst of \gx \citep{buxton13a, pawar13a}. We triggered observations with the \textsl{Nuclear Spectroscopic Telescope Array} \citep[\nustar,][]{harrison13a}, \swift \citep{swiftref}, and with the Australia Telescope Compact Array (ATCA). Overall we obtained five observations with \nustar, four during the rise of the outburst and one at the end of the outburst and \swift observations every other day, as well as three ATCA observations.
Figure~\ref{fig:batlc} shows the lightcurve of the outburst, as seen with the X-ray monitors \swift/BAT \citep{swiftbatref} and MAXI \citep{maxiref}. In a full outburst, the soft X-rays (as observed by MAXI) are expected to brighten as soon as the hard X-rays (as observed by BAT) decline, indicating the switch to the high/soft state. This outburst, however, did not follow that pattern and \gx never left the low/hard state resulting in a so-called failed outburst.
Table~\ref{tab:obsdates} gives a detailed observation log of the \nustar and simultaneous \swift observations.

\section{Observations and data reduction}
\label{sec:data}

\begin{deluxetable*}{lccccc}
\centering
\tablecolumns{8}
\tabletypesize{\scriptsize}
\tablecaption{\nustar and \swift observations and exposure times. \label{tab:obsdates}}
\tablehead{ \colhead{No.}  & \colhead{ObsID \nustar} & \colhead{ObsID \swift} &  \colhead{MJD range} & \colhead{exp. \nustar} & \colhead{exp. XRT}  \\
&  \colhead{ (800010130XX )} &   &  &\colhead{ FPMA / B [ks]} & \colhead{[ks]} }
\startdata 
I& 02 & 00032490015  & 56515.907--56516.994 &  42.25/ 42.26 & 1.1 \\
II & 04 & 00080180001 & 56520.709--56521.914 &  47.38 / 47.50  & 1.9 \\
III & 06 & 00080180002 and  & 56528.525--56529.792 &  43.78 / 43.94 & 1.6 \\
 & &  00032898009 & & & \\ 
IV & 08  & 00032898013 and & 56538.414--56539.897 & 61.94 / 62.29 & 2.0 \\
& & 00032898015 & & & \\
V & 10  & 00032988001 & 56581.994--56584.299 & 98.21 / 98.22 & 9.6 \\\hline
\swift IIIb & N/A & 00032898008 & 56527.788 & N/A & 0.97 \\
\swift IVb & N/A & 00032898024 & 56559.747 & N/A & 1.26 
\enddata
\end{deluxetable*}

\subsection{\swift}
\swift monitored the complete outburst with $\sim$1\,ks snapshots every other day. 
The evolution of the 2--10\,keV count rate 
(scaled to mCrab units) as measured by the \swift X-ray Telescope 
\citep[XRT,][]{swiftxrtref} is shown in Figure~\ref{fig:batlc}.  For data reduction, we used HEASOFT v6.15.1
and the XRT calibration released on 2013 March 13.  For each ObsID that
we used for the spectral analysis, we reprocessed the raw XRT data to produce
new event files using \texttt{xrtpipeline}.  Then, we used 
\texttt{xselect} to create source and background spectra.  We included
source counts from within 20 pixels ($47^{\prime\prime}$) of the \gx
position and background counts from an annulus centered on \gx
with an inner radius of 90 pixels and an outer radius of 110 pixels.
XRT was in Windowed Timing mode for the observations, and we scaled the
background to account for the active detector area.  For the response 
matrix, we used the file swxwt0to2s6\_20010101v015.rmf and \texttt{xrtmkarf}
to account for the effective area, including a correction using the 
exposure map for each observation.  The XRT spectra were rebinned to a 
\snr of 6 between 0.8--10\,keV using the Interactive Spectral Interpretation 
System \citep[ISIS,][]{houck00a}.  All analysis was done with ISIS 
v1.6.2-17 in this paper unless otherwise noted and uncertainties are given at the 90\% level. 
Observations \swift IIIb and IVb listed in Table~\ref{tab:obsdates} were not used in the X-ray spectral analysis as no simultaneous  \nustar data are available. We use those observations only for comparison with the radio flux, as they occurred closest in time to the radio observations described in Section~\ref{susec:atca}.

\subsection{\nustar}
\nustar consists of two independent grazing incidence telescopes which focus X-rays between 3--78\,keV onto corresponding focal planes consisting of cadmium zinc telluride (CZT) pixel detectors. \nustar, sensitive to X-ray energies from 3--79\,keV, provides unprecedented sensitivity and high spectral resolution at energies above 10\,keV, ideally suited to study the Compton reflection hump. The two focal planes are referred to as focal plane module (FPM) A and B. \nustar data were extracted using the standard \texttt{NUSTARDAS} v1.3.1 software. Source spectra were taken from a $120''$ radius region centered on the J2000 coordinates. The background was extracted as far away from the source as possible, from a $135''$ radius region. This approach induces small systematic uncertainties in the background, as the background is known to change over the field of view and from chip to chip \citep{wik14a}. However, \gx is  over five times brighter than the background even at the highest energies, so that these uncertainties are negligible. \nustar data were binned to a \snr of 36 in the relevant energy range of 4--78\,keV within ISIS. 
To reduce the spectral overlap which might be influenced by cross-calibration differences between \swift/XRT and \nustar, we exclude \nustar data below 4\,keV.

Timing analysis of \nustar observations I--IV is presented by \citet{bachetti15a}. The power spectral density (PSD) between 0.001--200\,Hz is consistent with typical hard-state PSDs and well described by three Lorentzian components \citep[see, e.g.][]{belloni05a}. We also calculated the PSD for observation V and find consistent results.

\subsection{ATCA}
\label{susec:atca}
We obtained quasi-simultaneous radio observations with ATCA, as a radio jet is expected to be launched during the low/hard state of \gx \citep{corbel00a}. 
ATCA's synthesis telescope is  located in Narrabri, New South Wales, Australia, and consists of six 22\,m antennas in  an east-west array, using 
 linearly orthogonal polarized feeds which allow the recording of full Stokes parameters.  The 
observations were conducted simultaneously at  5.5 and 9\,GHz on MJD 56516.4, 56527.46, and 56559.29, i.e. simultaneous to observation I, \swift IIIb, and \swift IVb (see Fig.~\ref{fig:batlc}), using the upgraded 
Compact Array Broadband Backend (CABB) system \citep{wilson11a}.  The first observation 
was conducted and reported by \citet{millerjones13a}. 
The array  was in a compact configuration  (H214 or H168) during  this period of ATCA observations. 

The amplitude and band-pass calibrator was PKS~1934$-$638, and the antennas' gain and phase 
calibration, as well as the polarization leakage, were usually derived from regular observations of the 
nearby  calibrator PMN~1646$-$50.  The editing, calibration, Fourier  transformation with multifrequency 
algorithms, deconvolution, and image analysis were performed using the {\tt MIRIAD} software package 
 \citep{sault98a}.  

\begin{figure}
\includegraphics[width=0.96\columnwidth]{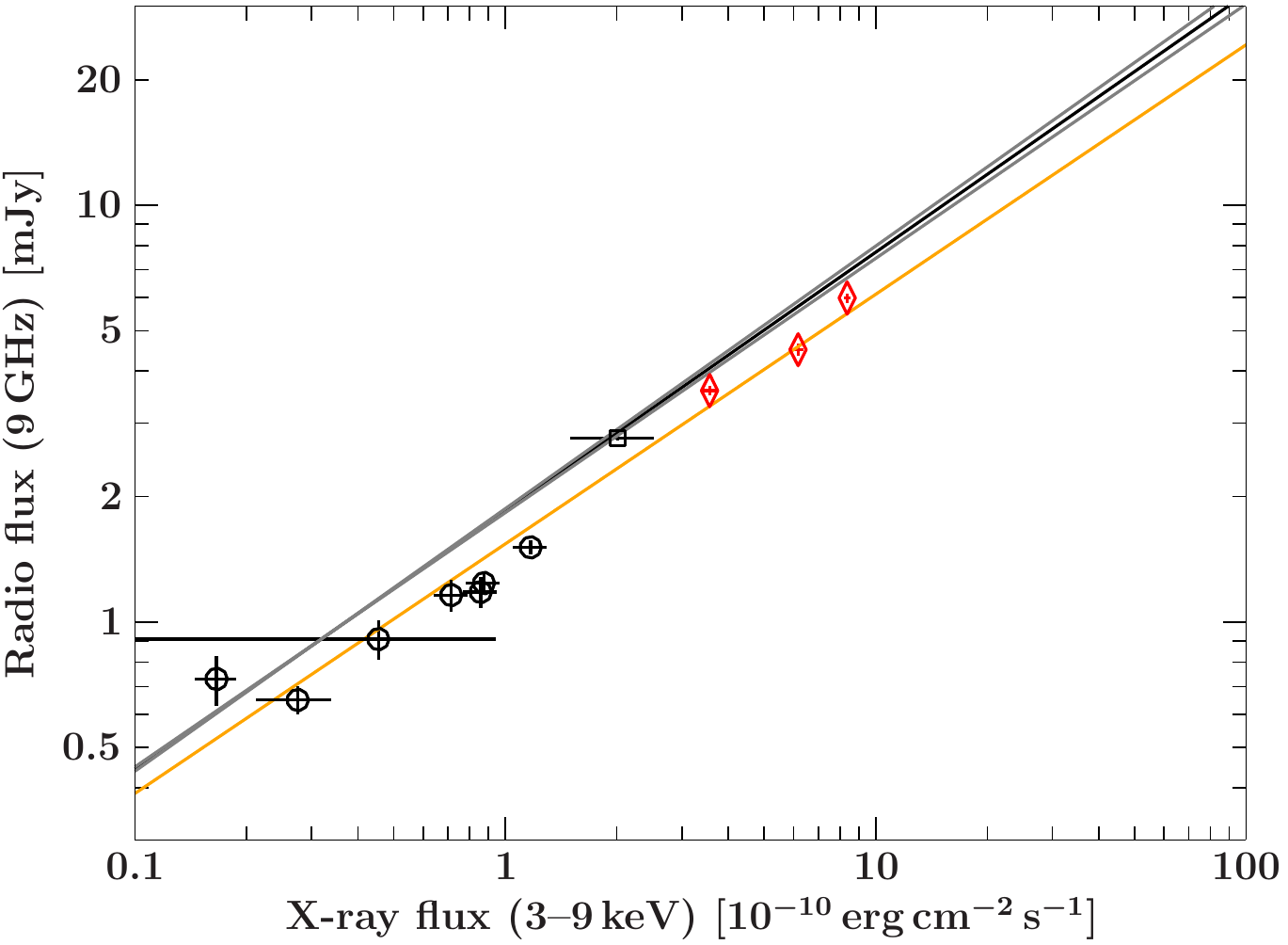} 
\caption{ATCA 9\,GHz flux as function of \swift/XRT 3--9\,keV flux.  Our 2013 data are shown with red diamonds, data from failed outbursts in 2008 and 2009 presented by \citet{corbel13a} as black circles and black squares, respectively. The superimposed black line is the best-fit correlation from all  archival data \citep[containing full and failed outbursts,][]{corbel13a}, the orange line is the best-fit to the observations during failed outbursts. Note that the error-bars on the 2013 data are smaller than the symbol size.}
\label{fig:radio}
\end{figure}

We show the observed 9\,GHz fluxes as a function of the 3--9\,keV X-ray flux as measured with \swift/XRT in Figure~\ref{fig:radio}  and compare them with archival data presented by \citet{corbel13a}. The data points fall below the measured correlation from all outbursts given by these authors and agree better when taking only the data of other failed outbursts in 2008 and 2009 into account. This behavior seems to indicate that the jet power is somewhat reduced in failed outbursts and hints at a different accretion geometry.

\section{Spectral Modeling}
\label{sec:spectra}

\subsection{Basic Fits}

\begin{figure}
\includegraphics[width=0.96\columnwidth]{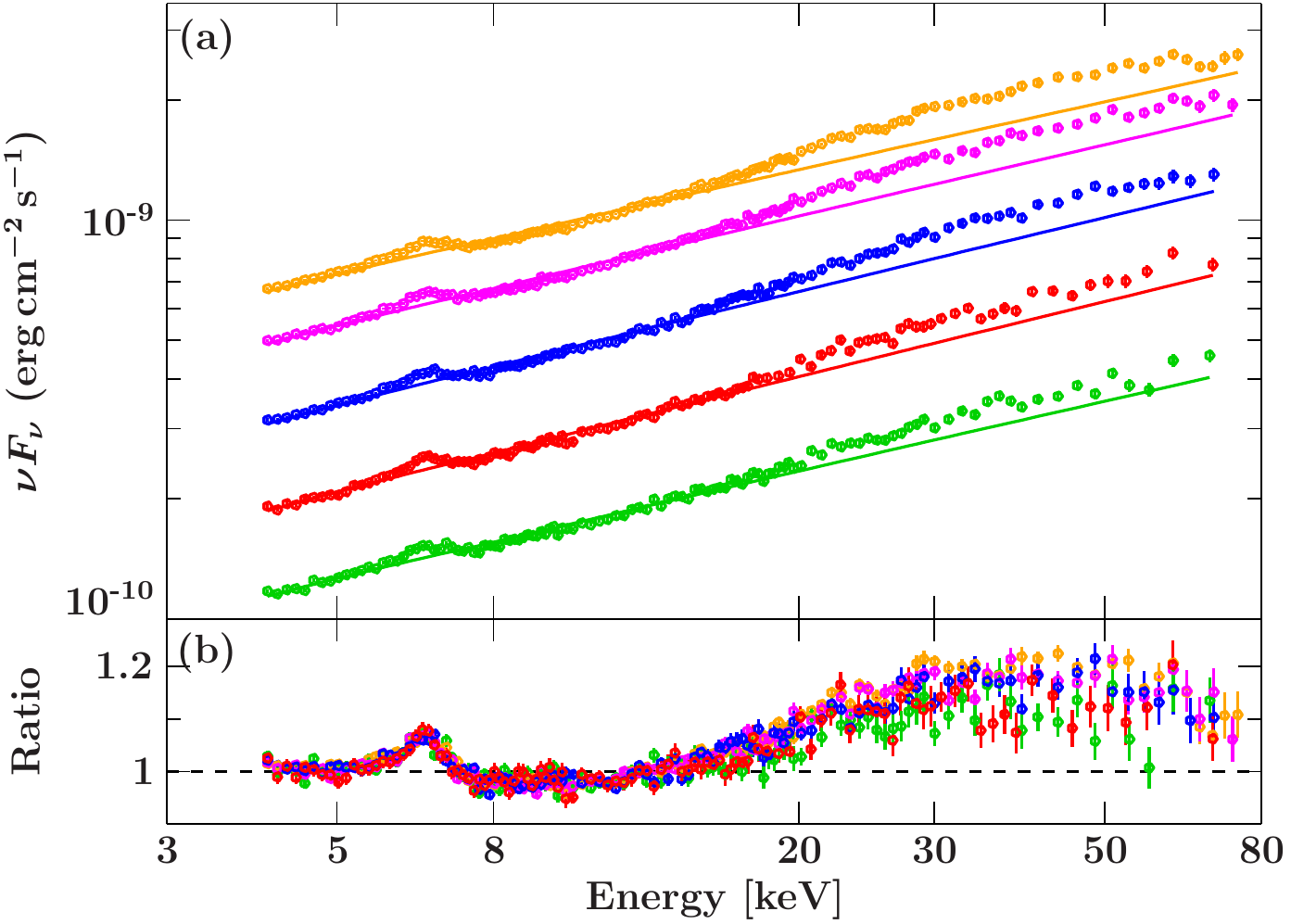} 
\caption{{(a)} Unfolded spectra of all five observations. Only \nustar/FPMA data are shown for clarity. Observation I is shown in red, observation II in blue, observation III in magenta, observation IV in orange and observation V in green. {(b)} Ratio residuals to a simple power-law fit.  See text for details.}
\label{fig:5obsunfold}
\end{figure}

As shown in Fig.~\ref{fig:5obsunfold} all five observations show very similar spectral shapes in \nustar, with clear evidence of reflection features. To highlight these features, we fit each \nustar observation with a simple absorbed power-law, using only data between 4--6\,keV, 8--10\,keV, and 50--78\,keV, i.e., ignoring the energy ranges where the strongest contribution from reflection features is expected. The residuals shown in Fig.~\ref{fig:5obsunfold} clearly indicate a strong \feka line and Compton hump. The shape of the iron lines appears to be constant over all observations, while we see indications that the Compton hump is more significant in the high flux data.

For the remainder of the paper we fit all \swift/XRT data between 0.8--10\,keV and all \nustar data between 4--78\,keV simultaneously, unless otherwise noted.
To model the reflection component we  add the \texttt{reflionx} model \citep{ross05a}, with its high-energy cutoff energy set to 500\,keV. While this model was originally calculated for AGN spectra, it has been applied successfully in the literature to describe black-hole binaries \citep[see, e.g., the discussion in][]{fabian12a}.
This model self-consistently describes the effects of reflection of an input power-law spectrum off an optically thick accretion disk, including the Compton hump and the fluorescent iron lines. For consistency we initially tie the photon index of the \texttt{reflionx} model to that of the primary power law (model M1).
To allow for relativistic effects close to the black hole we fold the model components through the \texttt{relconv} convolution kernel \citep{dauser10a}. We fix the outer radius of the accretion disk to $R_\text{out}= 400\,r_g$, as at these distances relativistic influences are negligible and the reflected flux is expected to be low.

In XSPEC notation the model is represented as \texttt{constant * tbabs * [powerlaw + relconv (reflionx)]}.
Here, \texttt{constant} takes into account small flux differences in the calibration of the different instruments. We quote all fluxes relative to \nustar/FPMA. 

While the continuum and reflection parameters are expected to change over the outburst, some parameters can be assumed to stay constant. In our model, these include the Galactic absorption towards the source (\nh), the black hole spin ($a$), the inclination of the accretion disk ($i$) and the iron abundance in the disk (Fe/solar), expressed in units of solar iron abundance. To obtain the best statistics for these parameters, we fit all five observations simultaneously, requiring that these parameters are the same for all observations.

The absorption is described using an updated version of the \texttt{tbabs} model \citep{wilms00a}, with the corresponding abundances and cross-sections from \citet{verner96a}. The fitted absorption column is only marginally higher than the expected Galactic absorption column of \mbox{$3.74\times10^{21}$\,cm$^{-2}$} \citep{kalberla05a} and is in good agreement with  values used in  literature for this source \citep[e.g.,][]{tomsick09a, plant14b}. We therefore do not add an additional absorption column intrinsic to the source.

As a first approach (model M1-q3, Table~\ref{tab:m1_q3}), we assume a standard Shakura-Sunyaev accretion disk \citep{shakura73a}, and we use an emissivity index ($q$) of 3 \citep{reynolds97b}.  The data quality does not allow us to constrain the spin of the black hole at the same time as the inner radius of the accretion disk. We therefore fix the spin to $a=0.92$, as measured by \citet{miller08a}. As described in Sect.~\ref{susec:spin}, the choice of $a$ does not influence our results significantly. 

Allowing the emissivity index $q$ to vary individually for each observation (model M1-qv, Table~\ref{tab:m1_qv}) improves the fit to $\redchi=1.30$ for 1698 dof and results in a steeper emissivity $\sim5$ for all observations but observation V, where it is fitted to $q_5=1.7\pm0.5$. All other parameters do not change significantly. 

The \texttt{relconv} model also allows us to use the ``lamppost'' geometry (M1-LP, Table~\ref{tab:m1_lp}), where the corona is assumed to be a point source on the spin axis above the black hole \citep{miniutti04a}. The main free parameter in this geometry is the height of the corona above the accretion disk plane and the illumination and emissivity index of the accretion disk is self-consistently calculated taking effects of general relativity into account \citep{dauser10a}. This model also provides a small improvement above the M1-q3 model ($\redchi=1.37$ for 1698 dof) and the coronal height is fitted to be $<4.2\,r_g$ for all observations. All other parameters do not change significantly.

Including a soft black-body component in any of the models does not improve the fit quality significantly nor does allowing the cutoff energy to vary. The 90\% lower limits on the cutoff energy are, in all observations, $>370$\,keV. Using models not accounting for a Compton hump, \citet{miyakawa08a} and \citet{motta09a} find somewhat lower cutoff energies at similar luminosities (0.9--5.6$\times10^{37}$\,erg\,s$^{-1}$ between 2--200\,keV). However, when including a reflection component,  \citet{miyakawa08a} do not find evidence for a cutoff at $E<500$\,keV at these luminosities. This is in agreement with results presented by \citet{plant14a}. The continuum in our data is therefore well described by a power-law, consistent with previous studies.

In all models (M1-q3, M1-qv, and M1-LP) we find an unphysically high iron abundance ($5.00^{+0.16}_{-0.12}$, $6.5\pm0.4$, and $5.27^{+0.37}_{-0.29}$ solar, respectively). In the literature, abundances of 1 or 2 times solar are typically assumed \citep{miller08a,reis08a,tamura12a,allured13a,plant14a}. Forcing a lower iron abundance results in clearly worse fits, with strong residuals around the \feka line energy. Using the \texttt{xillver} reflection model \citep{garcia10a} instead of \texttt{reflionx} does not improve the fit and requires an equally high or higher iron abundance.

\subsection{More sophisticated modeling}
\label{susec:bettermodels}

\subsubsection{Reflector sees a different continuum}
\label{sususec:2gamma}

The large iron abundance can be lowered and the statistically unacceptable \redchi values  improved by using more complex models, going beyond the traditional power-law plus reflection model. The standard geometries assume that the corona is a point-source and uniform, but we have to expect that the geometry in reality is more complex. The simplest approach to describe a corona which is physically extended with a non-uniform temperature profile is to allow different photon indices for the observed power law continuum and the input spectrum to the \texttt{reflionx} reflection (model M2). 

\begin{figure}
\includegraphics[width=0.96\columnwidth]{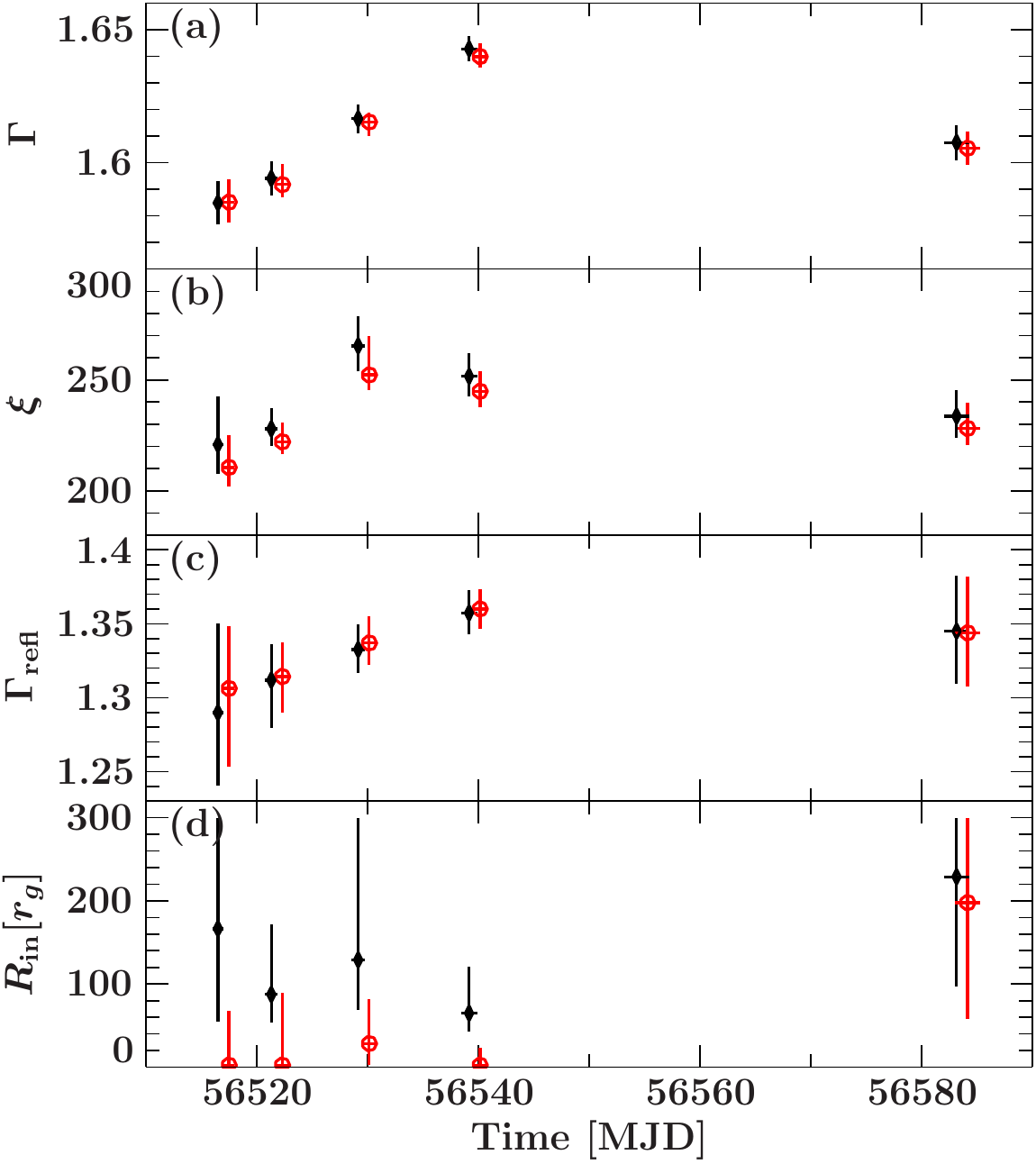} 
\caption{Results of the spectral fit for models M2-q3 and M2-LP.  Results of model M2-q3, i.e. with a fixed emissivity $q=3$, are shown as black diamonds, while results for model M2-LP, i.e. using the lamppost geometry, are shown as red circles. The latter are shifted in $x$-direction for clarity. \textit{(a)} photon index of the primary continuum, \textit{(b)} ionization parameter, \textit{(c)} input photon index for the reflection component, \textit{(d)} inner disk radius in $r_g$.}
\label{fig:resfit}
\end{figure}

When fixing the emissivity index to $q=3$ (M2-q3), this model improves the $\chi^2$ value significantly by $\Delta\chi^2=268$ for five additional parameters over M1-q3.
 We show the best-fit parameters for this model in Table~\ref{tab:q3_2gam}. We also apply the lamppost geometry (M2-LP) for which we give the best-fit values in Table~\ref{tab:lp_2gam}. Both models give mostly similar results and a comparable quality of the fit. The evolution of the spectral parameters with time for both models is shown in Figure~\ref{fig:resfit}. 

\begin{deluxetable*}{rlllll}
\tablewidth{0pc}
\tablecaption{Best-fit parameters for emissivity index $q=3$, spin $a=0.92$ and allowing for two different photon indices (M2-q3).\label{tab:q3_2gam}}
\tablehead{\colhead{Parameter} & \colhead{I} & \colhead{II} & \colhead{III}& \colhead{IV}& \colhead{V}}
\startdata
 $ N_\text{H}~[10^{22}\,\text{cm}^{-2}]$ & $0.868\pm0.020$ & --- & --- & --- & --- \\
 $ \text{Fe/solar}$ & $1.73^{+0.09}_{-0.08}$ & --- & --- & --- & --- \\
 $ i~[\text{deg}]$ & $48^{+12}_{-7}$ & --- & --- & --- & --- \\
\hline $ A_\text{cont}\tablenotemark{a}$ & $0.0669^{+0.0010}_{-0.0013}$ & $0.1112\pm0.0014$ & $0.1812\pm0.0019$ & $0.2538\pm0.0023$ & $0.0420\pm0.0006$ \\
 $ \Gamma_\text{power}$ & $1.585\pm0.009$ & $1.594\pm0.007$ & $1.616\pm0.006$ & $1.643\pm0.005$ & $1.608\pm0.007$ \\
 $ \Gamma_\text{refl}$ & $1.29^{+0.07}_{-0.05}$ & $1.312^{+0.025}_{-0.033}$ & $1.333^{+0.017}_{-0.016}$ & $1.357^{+0.016}_{-0.015}$ & $1.34\pm0.04$ \\
 $ A_\text{refl}\tablenotemark{a}$ & $\left(0.97^{+0.13}_{-0.15}\right)\times10^{-5}$ & $\left(1.67^{+0.12}_{-0.10}\right)\times10^{-5}$ & $\left(2.25\pm0.14\right)\times10^{-5}$ & $\left(3.24^{+0.16}_{-0.17}\right)\times10^{-5}$ & $\left(4.2\pm0.4\right)\times10^{-6}$ \\
 $ \xi$ & $\left(2.21^{+0.22}_{-0.14}\right)\times10^{2}$ & $\left(2.28^{+0.10}_{-0.08}\right)\times10^{2}$ & $\left(2.65^{+0.14}_{-0.12}\right)\times10^{2}$ & $\left(2.52^{+0.10}_{-0.09}\right)\times10^{2}$ & $\left(2.34^{+0.12}_{-0.10}\right)\times10^{2}$ \\
 $ R_\text{in}~[r_g]$ & $\left(1.7^{+1.4}_{-1.2}\right)\times10^{2}$ & $\left(0.9^{+0.9}_{-0.4}\right)\times10^{2}$ & $\left(1.3^{+1.8}_{-0.7}\right)\times10^{2}$ & $\left(0.65^{+0.57}_{-0.22}\right)\times10^{2}$ & $\left(2.3^{+0.8}_{-1.4}\right)\times10^{2}$ \\
 $ CC_\text{XRT}$ & $0.945\pm0.028$ & $0.915\pm0.015$ & $1.129\pm0.014$ & $1.104\pm0.012$ & $1.052\pm0.014$ \\
\hline $ \Delta \Gamma$ & $0.30\pm0.06$ & $0.282\pm0.029$ & $0.284\pm0.018$ & $0.285\pm0.016$ & $0.26\pm0.04$ \\
 $ R$ & 0.64 & 0.71 & 0.78 & 0.83 & 0.52 \\
 $ \%L_\text{edd}\tablenotemark{b}$ & 1.65 & 2.74 & 4.35 & 5.63 & 0.93 \\
\hline$\chi^2/\text{d.o.f.}$   & 2124/1698\\$\chi^2_\text{red}$   & 1.25\enddata
\tablenotetext{a}{in ph\,s$^{-1}$\,cm$^{-2}$}
\tablenotetext{b}{Luminosity calculated between 0.1--300\,keV assuming a distance of 8\,kpc and a black hole mass of 10\,$\text{M}_{\odot}$}
\end{deluxetable*}

\begin{deluxetable*}{rlllll}
\tablewidth{0pc}
\tablecaption{Same as Table~\ref{tab:q3_2gam} but for the lamppost geometry (M2-LP).\label{tab:lp_2gam}}
\tablehead{\colhead{Parameter} & \colhead{I} & \colhead{II} & \colhead{III}& \colhead{IV}& \colhead{V}}
\startdata
 $ N_\text{H}~[10^{22}\,\text{cm}^{-2}]$ & $0.851^{+0.020}_{-0.013}$ & --- & --- & --- & --- \\
 $ \text{Fe/solar}$ & $1.79^{+0.06}_{-0.08}$ & --- & --- & --- & --- \\
 $ i~[\text{deg}]$ & $49^{+7}_{-5}$ & --- & --- & --- & --- \\
\hline $ A_\text{cont}\tablenotemark{a}$ & $0.0668^{+0.0010}_{-0.0012}$ & $0.1108^{+0.0013}_{-0.0012}$ & $0.1809^{+0.0016}_{-0.0020}$ & $0.2529^{+0.0024}_{-0.0017}$ & $0.0418^{+0.0007}_{-0.0005}$ \\
 $ \Gamma_\text{power}$ & $1.585^{+0.008}_{-0.006}$ & $1.592^{+0.007}_{-0.004}$ & $1.615^{+0.004}_{-0.006}$ & $1.6402^{+0.0055}_{-0.0013}$ & $1.605\pm0.006$ \\
 $ \Gamma_\text{refl}$ & $1.31^{+0.05}_{-0.06}$ & $1.314^{+0.025}_{-0.028}$ & $1.336\pm0.015$ & $1.363^{+0.016}_{-0.013}$ & $1.35\pm0.04$ \\
 $ A_\text{refl}\tablenotemark{a}$ & $\left(1.02^{+0.08}_{-0.14}\right)\times10^{-5}$ & $\left(1.71\pm0.10\right)\times10^{-5}$ & $\left(2.36^{+0.10}_{-0.14}\right)\times10^{-5}$ & $\left(3.32^{+0.16}_{-0.08}\right)\times10^{-5}$ & $\left(4.2\pm0.4\right)\times10^{-6}$ \\
 $ \xi$ & $\left(2.11^{+0.14}_{-0.10}\right)\times10^{2}$ & $\left(2.23^{+0.08}_{-0.06}\right)\times10^{2}$ & $\left(2.54^{+0.15}_{-0.08}\right)\times10^{2}$ & $\left(2.45^{+0.09}_{-0.08}\right)\times10^{2}$ & $\left(2.29^{+0.10}_{-0.09}\right)\times10^{2}$ \\
 $ R_\text{in}~[r_g]$\tablenotemark{c} & $<66$ & $<74$ & $<59$ & $<26$ & $<300$ \\
 $ H \text{(corona)}~[r_g]$ & $\left(1.5^{+1.5}_{-0.7}\right)\times10^{2}$ & $\left(1.4^{+1.0}_{-0.6}\right)\times10^{2}$ & $\left(1.9^{+1.0}_{-0.7}\right)\times10^{2}$ & $\left(1.05^{+0.27}_{-0.39}\right)\times10^{2}$ & $\left(3.00^{+0.00}_{-2.98}\right)\times10^{2}$ \\
 $ CC_\text{XRT}$ & $0.946^{+0.025}_{-0.028}$ & $0.914^{+0.012}_{-0.014}$ & $1.128^{+0.012}_{-0.015}$ & $1.103\pm0.010$ & $1.050^{+0.010}_{-0.012}$ \\
\hline $ \Delta \Gamma$ & $0.28\pm0.05$ & $0.278\pm0.027$ & $0.278\pm0.016$ & $0.278\pm0.015$ & $0.26\pm0.04$ \\
 $ R$ & 0.63 & 0.70 & 0.77 & 0.81 & 0.51 \\
 $ L_x/L_\text{edd}\times100\tablenotemark{b}$ & 1.63 & 2.73 & 4.32 & 5.60 & 0.93 \\
\hline$\chi^2/\text{d.o.f.}$   & 2103/1693\\$\chi^2_\text{red}$   & 1.24\enddata
\tablenotetext{a}{in ph\,s$^{-1}$\,cm$^{-2}$}
\tablenotetext{b}{Luminosity calculated between 0.1--300\,keV assuming a distance of 8\,kpc and a black hole mass of 10\,$\text{M}_{\odot}$}
\tablenotetext{c}{The lower limit of $R_\text{in}$ is the ISCO at $2.2\,r_g$ for an assumed spin of $a=0.92$.}
\end{deluxetable*}

\begin{figure}
\includegraphics[width=0.96\columnwidth]{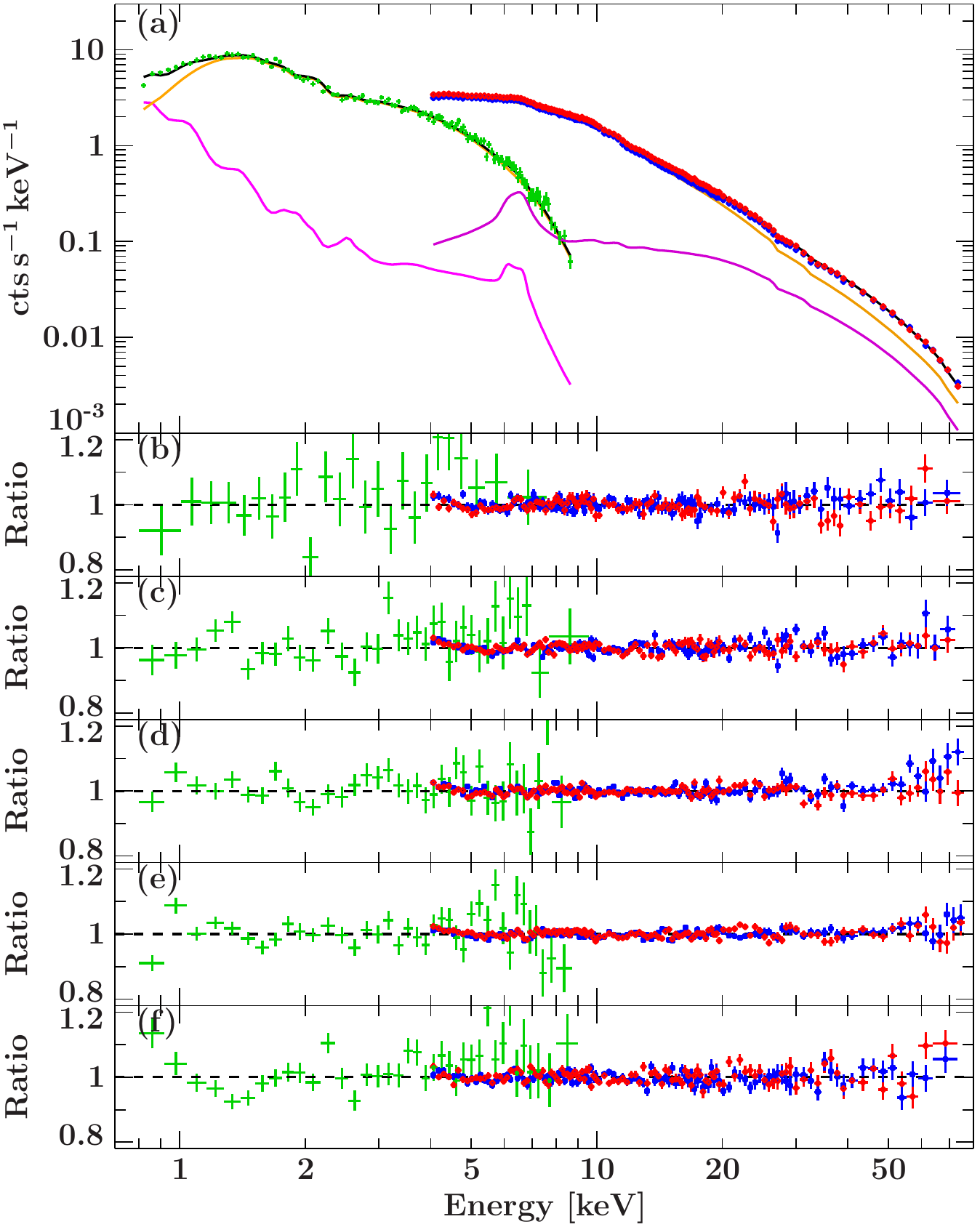} 
\caption{\textit{(a)} Data and best-fit model for M2-q3 for observation IV, showing the reflection component in magenta and the power-law components in orange. \swift/XRT data are shown in green, \nustar/FPMA in red, and FPMB in blue. \textit{(b)-(f)} Residuals in terms of model to data ratio of the best-fit model for  observation I--V, respectively. Data were rebinned for clarity. See Table~\ref{tab:q3_2gam} for the model parameters.}
\label{fig:5obs_residuals}
\end{figure}

We show the residuals for the M2-q3 model for all five observations in Figure~\ref{fig:5obs_residuals}. As can be seen, there is good agreement between all three instruments, though the \nustar data clearly dominate the statistics. Small systematic residuals at the lowest \nustar energy end can be attributed to calibration uncertainties. All five observations show very similar residuals and have a comparable quality of the fit.

The inner radius is highly unconstrained, with best fit values around $100\,r_g$ for M2-q3 but close to the ISCO for M2-LP. Both models are marginally consistent with values around $50\,r_g$. Due to the large uncertainties and geometry dependence of the values, it is not clear if a truncation of the accretion disk is present or not.
In the lamppost geometry the large uncertainties can be understood as being due to the large coronal height, which results in an emissivity index clearly below 3 between $\approx$3--100\,$r_g$ \citep{vaughan04a,dauser10a}. Therefore, the inner parts of the disk contribute less to the reflection model and the inner radius cannot be well constrained. 

We find very similar coronal heights around $150\,r_g$ for each observation with large uncertainties. The height is completely unconstrained in observation V in the allowed range between 3--300\,$r_g$. We therefore conclude that we see no indication for an evolution of the corona over the outburst and that the corona seems to be located relatively far away from the black hole. 

In the M2-q3 model the inclination is fitted to be $i=48^{\circ\,+12}_{~~-7}$ ($49^{\circ\,+7}_{~~-5}$ in M2-LP). These values are in very good agreement with the limits on the orbital inclination so that no misalignment between the accretion disk and the orbital plane is necessary.

We also obtain similar results when allowing the emissivity index to vary (M2-qv). The best-fit emissivity indices are relatively flat, around 1.6--1.8 and the models indicate a non-truncated accretion disk. The other parameters do not change significantly. The fit statistics improve by $\Delta \chi^2 = 46$ for five additional degrees of freedom over the M2-q3 model, which corresponds to an $F$-test false alarm probability of $6\times10^{-7}$, see Table~\ref{tab:m2_qv}.

The reflection fraction $R$ given in Tables~\ref{tab:q3_2gam}  and \ref{tab:lp_2gam} is the ratio of the flux of the reflection component between 0.001--1000\,keV to the flux of the primary unabsorbed power-law component between 0.1--1000\,keV. While this implies extrapolation of the model far outside the fitting range, it captures the whole energy range used in the calculation of \texttt{reflionx}. The reflection fraction is correlated with the X-ray luminosity, as already indicated by the variable Compton hump strength shown in Figure~\ref{fig:5obsunfold}. As the inner accretion disk radius seems to be constant, this indicates a change in coronal geometry where at higher flux more of the coronal emission is intercepted and reprocessed by the accretion disk.
 
The reflection fractions given in the tables are not directly comparable to the one given by \citet{plant14a}, as these authors use only the 4--10\,keV energy band to calculate it for the outbursts between 2002--2008. In this energy band we measure a reflection fraction around 0.05, about a factor of 2--3 lower than \citet{plant14a}.

The photon index incident to the reflection component is very hard ($\Gamma_\text{refl}\approx1.3$), indicating a strongly photon starved Comptonization region \citep{haardt91a}. Its difference to the photon index of the primary power law, $\Delta \Gamma$, is relatively constant over the outburst in all models; $\Delta \Gamma \approx 0.3$. 
 If this difference is due to the geometry of the corona, the corona seems to stay stable, even though the reflection fraction changes as a function of flux. We note, however, that the lamppost intrinsically assumes a point-like corona so that the different photon-indices we find here cannot directly be geometrically interpreted.

If we want to model an extended  corona along the spin axis, perhaps corresponding to the extended base of a jet \citep{markoff05a}, this can be done
more self-consistently in XSPEC or ISIS by adding a second relativistically smeared reflection component, with a coronal height larger than the first component and with its photon index tied to the observed primary continuum (M2b-LP). That is, we require that the observed continuum is also reflected and add as additional free parameters the flux of its reflection component and the second lamppost height. The incident spectrum to the first reflector is understood to originate much closer to the black hole such that it is invisible to the observer. This model results in a good fit, with \redchi=1.214 which corresponds to an improvement $\Delta \chi^2=54$ for six additional degrees of freedom over the M2-LP model. While this is only a marginal improvement, the added self-consistency is important for the physical interpretation of the model.

\begin{deluxetable*}{rlllll}
\tablewidth{0pc}
\tablecaption{Same as Table~\ref{tab:q3_2gam} but for the lamppost geometry and two reflection components whose input spectra originate from different heights (M2b-LP)\label{tab:lp_2gam2LP}}
\tablehead{\colhead{Parameter} & \colhead{I} & \colhead{II} & \colhead{III}& \colhead{IV}& \colhead{V}}
\startdata
 $ N_\text{H}~[10^{22}\,\text{cm}^{-2}] $ & $0.818\pm0.016$ & --- & --- & --- & --- \\
 $ \text{Fe/solar}$ & $2.56^{+0.31}_{-0.27}$ & --- & --- & --- & --- \\
 $ H_1 \text{(bottom)}~[r_g]$ & $2.2^{+1.0}_{-0.0}$ & --- & --- & --- & --- \\
 $ H_2 \text{(top)}~[r_g]$ & $\left(3.00^{+0.00}_{-1.44}\right)\times10^{2}$ & --- & --- & --- & --- \\
 $ i~[\text{deg}]$ & $39.2^{+2.7}_{-2.8}$ & --- & --- & --- & --- \\
\hline $ A_{\text{cont}}\tablenotemark{a}$ & $0.0643^{+0.0012}_{-0.0010}$ & $0.1075\pm0.0014$ & $0.1755^{+0.0018}_{-0.0017}$ & $0.2461^{+0.0020}_{-0.0018}$ & $0.0407\pm0.0007$ \\
 $ \Gamma_\text{power}$ & $1.568^{+0.010}_{-0.009}$ & $1.580^{+0.008}_{-0.007}$ & $1.602\pm0.006$ & $1.630\pm0.005$ & $1.590\pm0.008$ \\
 $ \Gamma_\text{refl}$ & $1.10^{+0.15}_{-0.10}$ & $1.19^{+0.08}_{-0.10}$ & $1.18^{+0.09}_{-0.17}$ & $1.16^{+0.08}_{-0.07}$ & $1.31^{+0.10}_{-0.31}$ \\
 $ A_{1,\text{refl}}\tablenotemark{a}$ & $\left(0.67^{+0.15}_{-0.13}\right)\times10^{-5}$ & $\left(1.34^{+0.16}_{-0.14}\right)\times10^{-5}$ & $\left(1.92^{+0.25}_{-0.18}\right)\times10^{-5}$ & $\left(2.80^{+0.19}_{-0.20}\right)\times10^{-5}$ & $\left(3.54^{+0.16}_{-1.29}\right)\times10^{-6}$ \\
 $ A_{2,\text{refl}}\tablenotemark{a}$ & $\left(4.026^{+0.010}_{-1.170}\right)\times10^{-6}$ & $\left(0.50^{+0.16}_{-0.14}\right)\times10^{-5}$ & $\left(0.62^{+0.27}_{-0.15}\right)\times10^{-5}$ & $\left(1.05\pm0.20\right)\times10^{-5}$ & $\left(0.6^{+2.0}_{-0.6}\right)\times10^{-6}$ \\
 $ \xi$ & $\left(2.13^{+0.15}_{-0.12}\right)\times10^{2}$ & $\left(2.20^{+0.10}_{-0.09}\right)\times10^{2}$ & $\left(2.59^{+0.20}_{-0.16}\right)\times10^{2}$ & $\left(2.429^{+0.148}_{-0.020}\right)\times10^{2}$ & $\left(2.26^{+0.14}_{-0.17}\right)\times10^{2}$ \\
 $ R_\text{in}~[r_g]$ & $6.3^{+10.7}_{-2.6}$ & $5.2^{+1.6}_{-1.4}$ & $4.3^{+1.5}_{-1.7}$ & $3.2\pm0.6$ & $\left(0.6^{+0.8}_{-0.6}\right)\times10^{2}$ \\
 $ CC_\text{XRT}$ & $0.941^{+0.028}_{-0.027}$ & $0.914\pm0.015$ & $1.131\pm0.014$ & $1.104\pm0.012$ & $1.053\pm0.014$ \\
\hline $ \Delta \Gamma$ & $0.47\pm0.12$ & $0.39\pm0.10$ & $0.43\pm0.13$ & $0.47\pm0.07$ & $0.28\pm0.20$ \\
 $ R$ & 0.63 & 0.69 & 0.77 & 0.84 & 0.47 \\
 $ \%L_\text{edd}\tablenotemark{b}$ & 1.65 & 2.73 & 4.34 & 5.62 & 0.94 \\
\hline$\chi^2/\text{d.o.f.}$   & 2053/1691\\$\chi^2_\text{red}$   & 1.21\enddata
\tablenotetext{a}{in ph\,s$^{-1}$\,cm$^{-2}$}
\tablenotetext{b}{Luminosity calculated between 0.1--300\,keV assuming a distance of 8\,kpc and a black hole mass of 10\,$\text{M}_{\odot}$}
\end{deluxetable*}

Table~\ref{tab:lp_2gam2LP} shows the best fit values of this model. As in the other M2 models, the inner accretion disk radius is consistent with being at the ISCO. The inner reflection height, $H_1$ is fit to the minimal allowed value, $2.2\,r_g$, while the outer reflection height, $H_2$, pegs at the upper limit, $300\,r_g$. Thus the model describes a very strongly elongated corona, and one would expect that intermediate heights also contribute to both the reflection and the observed primary continuum. Such a model, however, is not uniquely definable in ISIS or XSPEC as all reflectors and continua would be degenerate. We therefore take the two reflectors presented here as the best approximation to an elongated corona with varying power-law emission.

\subsubsection{Secondary continuum component}
\label{sususec:2cont}

In continuation of the idea that two different hard power-laws are present in the system, perhaps from gradients in the coronal temperature, we apply a model which consists of two power-law components, with only one being the input to the reflection component (M3). We tie the photon index of one power-law to the photon index of the reflection component. With respect to the models presented in the previous section (M2), we have another free parameter, the normalization of the second power-law component. 

\begin{deluxetable*}{rlllll}
\tablewidth{0pc}
\tablecaption{Same as Table~\ref{tab:q3_2gam} but for 2 power-law continua (M3-q3).\label{tab:q3_2cont}}
\tablehead{\colhead{Parameter} & \colhead{I} & \colhead{II} & \colhead{III}& \colhead{IV}& \colhead{V}}
\startdata
 $ N_\text{H}~[10^{22}\,\text{cm}^{-2}]$ & $0.904^{+0.022}_{-0.017}$ & --- & --- & --- & --- \\
 $ \text{Fe/solar}$ & $2.07^{+0.22}_{-0.19}$ & --- & --- & --- & --- \\
 $ i~[\text{deg}]$ & $31^{+6}_{-5}$ & --- & --- & --- & --- \\
\hline $ A_\text{cont,1}\tablenotemark{a}$ & $0.034^{+0.009}_{-0.014}$ & $0.054^{+0.010}_{-0.019}$ & $0.075^{+0.015}_{-0.020}$ & $0.095^{+0.021}_{-0.031}$ & $0.00027^{+0.01137}_{-0.00027}$ \\
 $ \Gamma_\text{cont,1 and refl}$ & $1.426^{+0.028}_{-0.039}$ & $1.420^{+0.021}_{-0.036}$ & $1.417^{+0.015}_{-0.019}$ & $1.433^{+0.015}_{-0.020}$ & $1.39^{+0.05}_{-0.04}$ \\
 $ A_\text{refl}$ & $\left(0.59^{+0.09}_{-0.08}\right)\times10^{-5}$ & $\left(1.11^{+0.16}_{-0.14}\right)\times10^{-5}$ & $\left(1.52^{+0.20}_{-0.19}\right)\times10^{-5}$ & $\left(2.3\pm0.4\right)\times10^{-5}$ & $\left(3.6^{+0.4}_{-0.7}\right)\times10^{-6}$ \\
 $ \xi$ & $\left(2.51^{+0.25}_{-0.20}\right)\times10^{2}$ & $\left(2.49^{+0.22}_{-0.15}\right)\times10^{2}$ & $\left(3.03^{+0.35}_{-0.22}\right)\times10^{2}$ & $\left(2.85^{+0.36}_{-0.21}\right)\times10^{2}$ & $\left(2.53^{+0.22}_{-0.16}\right)\times10^{2}$ \\
 $ R_\text{in}~[r_g]$ & $\left(1.55^{+0.00}_{-0.98}\right)\times10^{2}$ & $44^{+29}_{-15}$ & $\left(0.52^{+0.63}_{-0.20}\right)\times10^{2}$ & $26^{+15}_{-7}$ & $\left(0.7^{+0.9}_{-0.4}\right)\times10^{2}$ \\
 $ A_\text{cont,2}\tablenotemark{a}$ & $0.037^{+0.010}_{-0.007}$ & $0.066^{+0.016}_{-0.008}$ & $0.119^{+0.016}_{-0.010}$ & $0.171^{+0.027}_{-0.016}$ & $0.0412^{+0.0009}_{-0.0106}$ \\
 $ \Gamma_\text{cont,2}$ & $1.91^{+0.24}_{-0.18}$ & $1.92^{+0.17}_{-0.14}$ & $1.89^{+0.13}_{-0.10}$ & $1.87^{+0.12}_{-0.09}$ & $1.603^{+0.105}_{-0.010}$ \\
 $ CC_\text{XRT}$ & $0.93\pm0.04$ & $0.900\pm0.018$ & $1.120\pm0.017$ & $1.100\pm0.013$ & $1.068^{+0.014}_{-0.017}$ \\
\hline $ \Delta \Gamma$ & $0.48\pm0.21$ & $0.50\pm0.16$ & $0.47\pm0.12$ & $0.44\pm0.10$ & $0.22\pm0.08$ \\
 $ R_1$ & 0.46 & 0.52 & 0.62 & 0.74 & 29.87 \\
 $ R_{1+2}$ & 0.31 & 0.33 & 0.36 & 0.39 & 0.49 \\
 $ \%L_\text{edd}\tablenotemark{b}$ & 2.16 & 3.66 & 5.88 & 7.71 & 0.94 \\
\hline$\chi^2/\text{d.o.f.}$   & 2072/1693\\$\chi^2_\text{red}$   & 1.22\enddata
\tablenotetext{a}{in ph\,s$^{-1}$\,cm$^{-2}$}
\tablenotetext{b}{Luminosity calculated between 0.1--300\,keV assuming a distance of 8\,kpc and a black hole mass of 10\,\msun}
\end{deluxetable*}

\begin{figure}
\includegraphics[width=0.96\columnwidth]{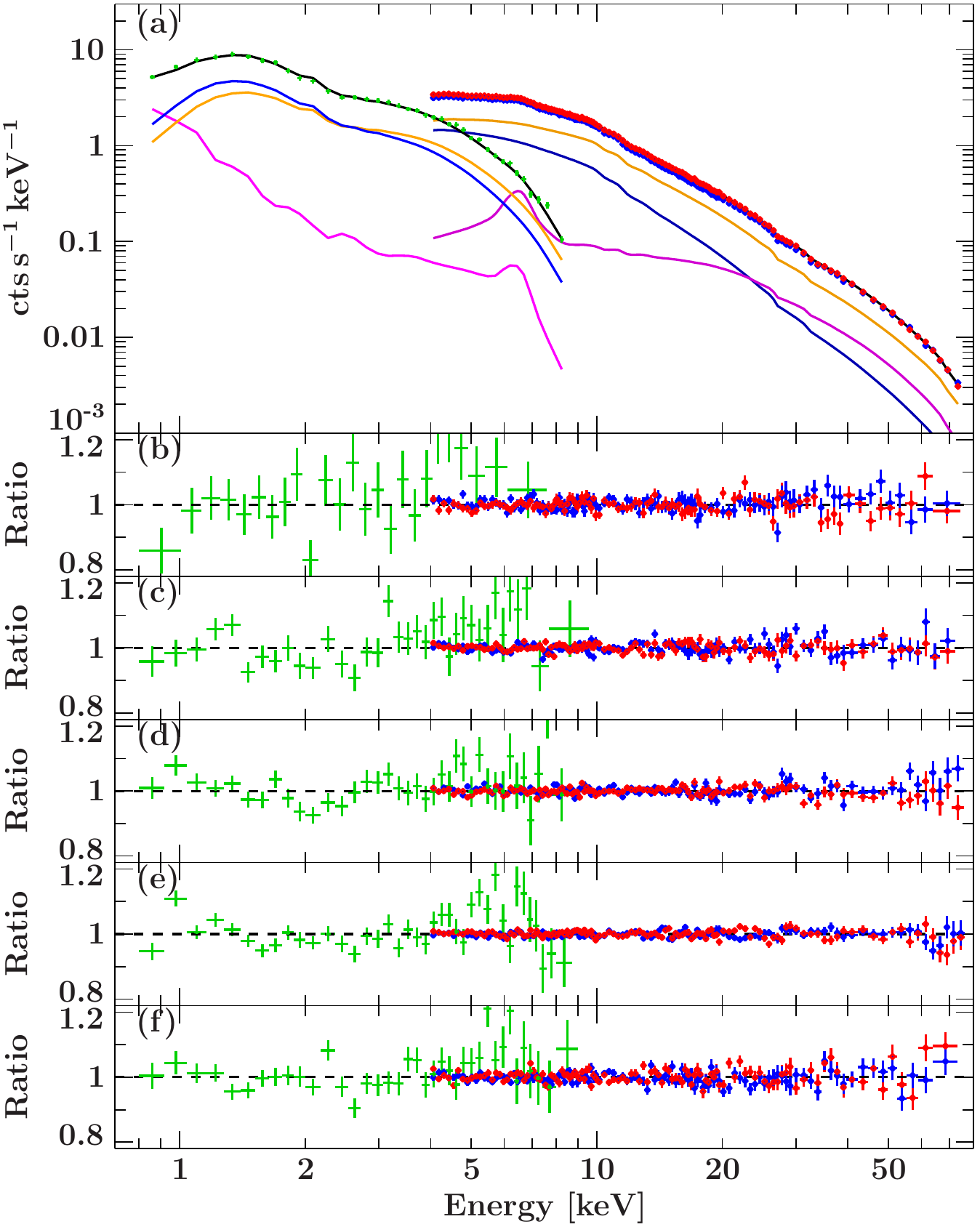} 
\caption{Same as Figure~\ref{fig:5obs_residuals} but for model M3-q3. The power-law incident to the reflection model is shown in orange, the second power-law in blue. Residuals are shown in units of $\sigma$. The best-fit parameters are given in Table~\ref{tab:q3_2cont}.}
\label{fig:5obs_res_2cont}
\end{figure}

We again fit three different models: the first with the emissivity fixed at $q=3$ (M3-q3), the second allowing the emissivity index to vary (M3-qv), and finally using the lamppost geometry (M3-LP). Figure~\ref{fig:5obs_res_2cont} shows the residuals for M3-q3, separately for each observation for clarity. Table~\ref{tab:q3_2cont} gives the best-fit values for M3-q3. 
We give two values for the reflection fraction: $R_1$ is calculated using only the first continuum, which has the photon index used for the reflection component. $R_{1+2}$ is calculated using the total observed flux from both continuum models.

All three models result in very comparable qualities of fit with \redchi/dof=1.22/1693~(M3-q3), 1.21/1688 (M3-qv), and 1.22/1686 (M3-LP). However, the relative strengths of the two continuum components depend on the assumed geometry. The lamppost model, in particular,  gives very different results (all model parameters can be found in Tables~\ref{tab:m3_qv} and \ref{tab:m3_lp}). Additionally, in observation V, the first power-law normalization  in all three models is consistent with $0$, reducing this model to the M2 described in Sect.~\ref{sususec:2gamma}.
In the M3-q3 model the second, non-reflected power-law contributes between 30--50\% of the broad-band flux in observations I--IV.

The best fit iron abundance is found to be between 2--2.5 the solar value, depending on the geometry. The inner radius of the accretion disk is again consistent with being  close to the ISCO, especially in the model with a free emissivity index where the best fit values are below 20\,$r_g$. The differences between the two photon-indices are typically larger than in the previous model, with values around $\Delta\Gamma=0.5$. 
Overall, this model confirms the previous results that we can obtain a reasonable iron abundance by  using a more complicated continuum model and that the accretion disk is not significantly truncated. However, through the introduction of another free parameter, the parameters are less well constrained and the fit quality is not significantly improved.

\subsection{Iron line complex}
\label{susect:iron}

\begin{deluxetable*}{rlllll}
\tablewidth{0pc}
\tablecaption{Same as Table~\ref{tab:q3_2gam} but adding an additional narrow Gaussian line at 6.4\,keV (M2-q3-Fe).\label{tab:eg_2gam}}
\tablehead{\colhead{Parameter} & \colhead{I} & \colhead{II} & \colhead{III}& \colhead{IV}& \colhead{V}}
\startdata
 $ N_\text{H}~[10^{22}\,\text{cm}^{-2}]$ & $0.853^{+0.020}_{-0.017}$ & --- & --- & --- & --- \\
 $ \text{Fe/solar}$ & $1.58^{+0.10}_{-0.09}$ & --- & --- & --- & --- \\
 $ i~[\text{deg}]$ & $59^{+17}_{-9}$ & --- & --- & --- & --- \\
\hline $ A_\text{cont}\tablenotemark{a}$ & $0.0667\pm0.0010$ & $0.1114\pm0.0014$ & $0.1823\pm0.0018$ & $0.2556\pm0.0023$ & $0.0418\pm0.0006$ \\
 $ \Gamma_\text{cont}$ & $1.583^{+0.010}_{-0.009}$ & $1.594\pm0.007$ & $1.6177^{+0.0054}_{-0.0023}$ & $1.645\pm0.005$ & $1.603^{+0.008}_{-0.004}$ \\
 $ \Gamma_\text{refl}$ & $1.24\pm0.06$ & $1.27\pm0.04$ & $1.280^{+0.026}_{-0.027}$ & $1.319^{+0.017}_{-0.016}$ & $1.25\pm0.06$ \\
 $ A_\text{refl}\tablenotemark{a}$ & $\left(1.08^{+1.07}_{-0.11}\right)\times10^{-5}$ & $\left(1.75\pm0.12\right)\times10^{-5}$ & $\left(2.40\pm0.16\right)\times10^{-5}$ & $\left(3.45^{+0.17}_{-0.18}\right)\times10^{-5}$ & $\left(4.4\pm0.5\right)\times10^{-6}$ \\
 $ \xi$ & $\left(2.10^{+0.13}_{-1.07}\right)\times10^{2}$ & $\left(2.24^{+0.09}_{-0.08}\right)\times10^{2}$ & $\left(2.58^{+0.14}_{-0.05}\right)\times10^{2}$ & $\left(2.44^{+0.10}_{-0.04}\right)\times10^{2}$ & $\left(2.31^{+0.13}_{-0.10}\right)\times10^{2}$ \\
 $ R_\text{in}~[r_g]$ & $39^{+34}_{-17}$ & $\left(0.72^{+0.63}_{-0.30}\right)\times10^{2}$ & $51^{+36}_{-21}$ & $40^{+20}_{-13}$ & $45^{+42}_{-21}$ \\
 $ A_{\text{FeK}\alpha}\tablenotemark{a}$ & $\left(0.77^{+0.24}_{-0.25}\right)\times10^{-4}$ & $\left(0.7\pm0.4\right)\times10^{-4}$ & $\left(1.7^{+0.4}_{-0.5}\right)\times10^{-4}$ & $\left(2.1\pm0.5\right)\times10^{-4}$ & $\left(0.51\pm0.13\right)\times10^{-4}$ \\
 $ CC_\text{XRT}$ & $0.949\pm0.028$ & $0.912\pm0.015$ & $1.124\pm0.014$ & $1.099\pm0.012$ & $1.052\pm0.014$ \\
\hline $ \Delta \Gamma$ & $0.34\pm0.06$ & $0.33\pm0.04$ & $0.338\pm0.027$ & $0.327\pm0.017$ & $0.36\pm0.06$ \\
 $ R$ & 0.65 & 0.73 & 0.80 & 0.85 & 0.53 \\
Eqw (FeK$\alpha$) [eV] & $22^{+7}_{-8}$ & $12\pm7$ & $19\pm5$ & $17\pm4$ & $24^{+6}_{-7}$ \\
 $ \%L_\text{edd}$ & 1.64 & 2.74 & 4.36 & 5.62 & 0.94 \\
\hline$\chi^2/\text{d.o.f.}$   & 2009/1693\\$\chi^2_\text{red}$   & 1.19\enddata
\tablenotetext{a}{in ph\,s$^{-1}$\,cm$^{-2}$}
\tablenotetext{b}{Luminosity calculated between 0.1--300\,keV assuming a distance of 8\,kpc and a black hole mass of 10\,$\text{M}_{\odot}$}
\end{deluxetable*}

In all previous fits we modeled the \feka line self-consistently as arising from  reflection off an accretion disk with constant ionization, as described by the \texttt{reflionx} model. However, close inspection of the residuals reveals that none of these models captures the shape of the line perfectly. We therefore add an ad-hoc narrow Gaussian component to the M2 models, with its energy fixed at 6.4\,keV, the energy of neutral \feka. This line could, for example, be produced further out in the disk, where we encounter near neutral iron and the influence of the relativistic effects of the black hole are negligible. With this addition, we find significantly better fits for all versions of the M2 model, i.e., when the photon indices between the continuum and reflector are independent.  

We note that \citet{plant14b} do not see evidence for such an additional narrow component. However, their data lack the crucial coverage above 10\,keV and therefore the reflection model is entirely driven by the iron line shape. When applying a model similar to their best-fit model using the lamppost geometry to our observation IV, and limiting the fitting range to below 10\,keV, we obtain very similar results, most notably an ionization around $\xi=1000$ and no residuals in the iron line band. However, this model clearly underpredicts the flux in the hard X-ray band, especially  around the Compton hump.

As an example of a model with an additional Gaussian component, we give the best values for a fixed emissivity index of $q=3$ in Table~\ref{tab:eg_2gam} (M2-q3-Fe). The narrow core is fitted to an equivalent width around 20\,eV in all observations and does not show a dependence on flux. Fits for a variable emissivity index (M2-qv-Fe) and the lamppost geometry (M2-LP-Fe), can be found in Tables~\ref{tab:m2_qv_fe} and \ref{tab:m2_lp_fe}, respectively. Note that a variable emissivity results in best-fit values consistent with $q=3$. 

As can be seen in Table~\ref{tab:eg_2gam}, the inclination is higher than in our other models, particularly compared to M2b-LP and M3-q3.
However, it is still consistent with being below $60^\circ$ as required by the lack of eclipses 
and consistent with all M2 models.  See Section~\ref{susec:spin} for a more detailed discussion about the inclination.  Furthermore, this is the only model where we find a significant truncation of the inner accretion disk around 30\,$r_g$. 
None of the other model parameters changes significantly.

\section{Discussion}
\label{sec:summ}
We have presented simultaneous spectral fits to five  \swift/XRT and \nustar observations during the failed 2013 outburst of \gx covering luminosities between $\sim$0.9--6\%\,\ledd. All observations show a very hard power-law, with a photon index $\Gamma \approx 1.6$ and clear evidence for reflection. We have shown that standard models, consisting of a power-law continuum and an additive reflection model fail to reproduce the data within sensible physical parameters. The data, thanks to the very high \snr, clearly indicate that the input to the reflection model needs to be different from the continuum emission observed directly. Our best fit solutions typically require that the inner radius of the accretion disk extends close to the ISCO ($R_\text{in} \ll 100\,r_g$). 

For comparison, we analyzed \xmm data presented by \citet{plant14b} taken during the same outburst (between our observations IV and V; \xmm ObsIDs 0692341201, 0692341301, and 0692341401). We used a similar  annular extraction  region as these authors with an inner radius of $11.5''$ and an outer radius of $45''$ to excise the heavily piled-up core. Applying the best-fit \texttt{relxill} model presented by \citet{plant14b} and fitting all three observations simultaneously results in an acceptable fit with \redchi/dof=1.25/5079. When applying model M2-q3, and fixing the secondary power-law and the ionization to the best-fit \nustar values, we find a very similar quality of fit with  \redchi/dof=1.26/5081 and similar structures in the residuals. This similarity shows that the soft X-ray band-pass of \xmm is not sufficient to constrain the complex accretion geometry, as the Compton hump is not covered and that at the same time our model is fully compatible with the \xmm data.

Besides the input power-law  to the reflector, the biggest difference between our models and the ones presented by \citet{plant14b} is that they find evidence for the presence of a cold disk thermal component. Such a component is often observed in \gx in the hard state  \citep[see, e.g.][]{miller06a, reis08a, wilkinson09a}, but we do not find an improvement by adding it.
Its absence is likely connected to the less sensitive XRT data compared to \xmm as well as a  degeneracy between the absorption column and the disk. This degeneracy is seen when describing the \xmm data of \citet{plant14b} with the M2-q3 model, where we find slightly lower values of the absorption column ($\nh=(0.664\pm0.006)\times10^{22}$\,cm$^{-2}$ compared to  $\approx(0.74\pm0.01)\times10^{22}$\,cm$^{-2}$).

We note that the calculated reflection fractions are all below unity. In a static geometry, the reflection fraction corresponds approximately to the solid angle covered by the reflector in units of $2\pi$, i.e., a reflection fraction of 1 would correspond to an infinite slab illuminated from above. We measure values below 0.5, which could indicate a truncation of the accretion disk at the inner radius. On the other hand, a outflowing corona would also result in a reduced reflection fraction, as the coronal radiation is beamed away from the accretion disk \citep{beloborodov99a}.

The photon index of the observed primary continuum seems to show a hysteresis effect, with observation V showing a significantly softer spectrum than observations I and II, despite being much fainter. 
This hysteresis could be connected to the hysteresis observed in the radio/X-ray and NIR/X-ray correlation \citep{corbel13a, russell07a} which might be related to different jet behavior between rising and decaying hard state observations. However, this hysteresis is typically observed in a full outburst, in which the hard states are separated by a soft or intermediate state, where we expect stronger changes in the accretion geometry.

With our complex corona models (M2, M2b, and M3), we find an iron abundance of typically $\sim$1.8 times solar, in line with previous work and as expected in LMXBs \citep[e.g.][]{allured13a}. The fit statistically constrains the iron abundance  extremely well, but the value strongly depends on the model assumptions. For example, with a variable emissivity index (models M2-qv and M3-qv) we find an iron abundance around 2.5 times solar, while it drops to around 1.5 solar in all models when a narrow Gaussian line is included (e.g., M2-q3-Fe).
The absolute value and the error bars are therefore model dependent and subject to systematic uncertainties not accounted for in the tables.

\subsection{Accretion geometry}

As discussed in \citet{fabian14a}, due to light-bending, the regions of the corona closer to the black hole  contribute more irradiating flux on the accretion disk than regions further away, while the opposite is true for the primary observed continuum. A change in the coronal parameters with height will therefore result in an observed continuum with a different spectral index than the one incident to the reflector. This effect is most relevant if the corona extends close to the black hole where relativistic effects are strongest \citep{dauser13a}.

We can best approximate this geometry  by using model M2b-LP consisting of two coronae at different heights above the black hole  with different photon indices, as presented in Table~\ref{tab:lp_2gam2LP}. We find that the inner corona originates very close to the black hole, and thus is subject to strong light bending effects. These effects prevent most of the flux from that part of the corona from reaching the observer directly, and we only  see the reflected part. 

We also find evidence for a complex shape of the \feka line, which is clearly relativistically broadened with an additional narrow core, close to 6.4\,keV.  A detailed discussion of its shape is beyond the scope of this paper and will be presented in a forthcoming publication making use of \suz/XIS data with better spectral resolution (Tomsick et al., in prep.). The broadened component is constrained in our fits through the \texttt{relconv} smearing kernel. 
In the models without an extra component for the narrow core, the presence of a reflector within 
$\ll$100\,$r_g$ is clearly required (see Table~\ref{tab:lp_2gam}). Adding the narrow core moves the lower limit of the inner accretion disk radius out to about 20\,$r_g$.

For the faintest observation (V) all models indicate the possibility that the accretion disk is truncated. These values are only weakly constrained and strongly dependent on the assumed X-ray source geometry (see Tables~\ref{tab:q3_2gam} and \ref{tab:lp_2gam}). It is clear, however, that during observation V the source spectrum was different from the first four observations (see also Figure~\ref{fig:resfit}) with either a change in the inner accretion disk radius or the corona's location and spectrum or a combination thereof.

\subsection{Inclination and spin}
\label{susec:spin}
All our models indicate an inclination in the range around $50^\circ$. This value is significantly higher than $i\approx20\text{--}30^\circ$ found when modeling other observations with similar reflection models \citep{miller04a,reis08a, plant14b}. 
\citet{kolehmainen10a} find that inclinations $i>45^\circ$ give better fits in the high-soft state when fitting the disk continuum to measure the spin. Our result seems to reconcile the continuum fits with the reflection fits with respect to the disk inclination. 
All of these previous data, however, were taken during a full outburst, i.e., an outburst that followed the standard evolution through the high-soft state. The data presented here were taken during a failed outburst, which might have a different geometry. 

Our models do not constrain the spin. We use $a=0.92$ throughout, as measured by \citet{miller04a}. However, as the inner radius is only weakly constrained and typically of the order of $10\,r_g$, lower spin values are completely consistent with our results. To test this, we set the spin to $a=0$ for the M2-q3 model and obtained basically identical fits (see Table~\ref{tab:m2_q3_a0}).

\section{Summary}
While the combined \nustar and \swift data provide one of the best data-sets on \gx in the low/hard state to date, we have shown that it is difficult to measure the inner truncation radius of the accretion disk precisely. The measured value depends strongly on the assumed geometry and emissivity profile of the accretion disk. However, we find no evidence for a strongly truncated disk, i.e., with an inner radius $>100\,r_g$. Furthermore, our spectral fits clearly show that the continuum spectrum incident to the reflector is significantly different from the observed primary continuum. The spectrum reflected by the accretion disk is significantly harder, which is necessary to explain the relative strength of the Compton hump to the \feka line. A lamppost geometry with changing spectral hardness as a function of coronal height seems to describe the observed spectra well, but can only be regarded as a crude approximation to the true physical geometry.

We would like to stress again that the data were taken during a failed outburst, during which the source did not switch into the soft state. It is currently unknown what the difference between failed and standard outbursts is, and how a transition to the soft state is triggered. Continued monitoring of \gx and similar black hole transients is necessary to answer these questions and study if we can measure significantly different accretion geometries in these two types of outbursts.

\acknowledgments
We thank the anonymous referee for the very constructive and helpful comments.
This work was supported under NASA Contract No. NNG08FD60C, and
made use of data from the {\it NuSTAR} mission, a project led by
the California Institute of Technology, managed by the Jet Propulsion
Laboratory, and funded by the National Aeronautics and Space
Administration. We thank the {\it NuSTAR} Operations, Software and
Calibration teams for support with the execution and analysis of these observations. This research has made use of the {\it NuSTAR}
Data Analysis Software (NuSTARDAS) jointly developed by the ASI
Science Data Center (ASDC, Italy) and the California Institute of
Technology (USA). 
JAT acknowledges partial support from NASA \swift Guest Investigator grant NNX13AJ81G and NNX14AC56G.
SC acknowledges funding support from the ANR ``CHAOS'' (ANR-12-BS05-0009).
We would like to thank John E. Davis for the \texttt{slxfig} module, which was used to produce all figures in this work. 

\begin{appendix}
\section{Additional model parameters}

\begin{deluxetable*}{rlllll}
\tablewidth{0pc}
\tablecaption{Best-fit parameters for an emissivity index $q=3$ and spin $a=0.92$ (M1-q3).\label{tab:m1_q3}}
\tablehead{\colhead{Parameter} & \colhead{I} & \colhead{II} & \colhead{III}& \colhead{IV}& \colhead{V}}
\startdata
 $ N_\text{H}~[10^{22}\,\text{cm}^{-2}]$ & $0.783^{+0.020}_{-0.016}$ & --- & --- & --- & --- \\
 $ \text{Fe/solar}$ & $5.00^{+0.16}_{-0.12}$ & --- & --- & --- & --- \\
 $ i~[\text{deg}]$ & $39.9^{+0.7}_{-1.2}$ & --- & --- & --- & --- \\
\hline $ A_\text{cont}\tablenotemark{a}$ & $0.0607\pm0.0007$ & $0.0999\pm0.0008$ & $0.1620^{+0.0010}_{-0.0011}$ & $0.2268^{+0.0012}_{-0.0014}$ & $0.0391\pm0.0004$ \\
 $ \Gamma$ & $1.540\pm0.007$ & $1.544\pm0.005$ & $1.562\pm0.004$ & $1.587\pm0.004$ & $1.573\pm0.006$ \\
 $ A_\text{refl}\tablenotemark{a}$ & $\left(0.97^{+0.21}_{-0.10}\right)\times10^{-5}$ & $\left(1.67^{+0.10}_{-0.11}\right)\times10^{-5}$ & $\left(2.41^{+0.13}_{-0.16}\right)\times10^{-5}$ & $\left(3.57^{+0.14}_{-0.19}\right)\times10^{-5}$ & $\left(4.6\pm0.4\right)\times10^{-6}$ \\
 $ \xi$ & $\left(2.04^{+0.12}_{-0.44}\right)\times10^{2}$ & $\left(2.11^{+0.08}_{-0.06}\right)\times10^{2}$ & $\left(2.32^{+0.11}_{-0.08}\right)\times10^{2}$ & $\left(2.20^{+0.07}_{-0.05}\right)\times10^{2}$ & $\left(2.09^{+0.08}_{-0.06}\right)\times10^{2}$ \\
 $ R_\text{in}~[r_g]$ & $4.6^{+0.9}_{-0.6}$ & $4.2^{+0.6}_{-0.5}$ & $4.0^{+0.5}_{-0.4}$ & $3.89^{+0.40}_{-0.26}$ & $4.4^{+0.8}_{-0.5}$ \\
 $ CC_\text{XRT}$ & $0.962^{+0.027}_{-0.028}$ & $0.936^{+0.014}_{-0.015}$ & $1.163^{+0.015}_{-0.014}$ & $1.138^{+0.013}_{-0.012}$ & $1.061^{+0.015}_{-0.014}$ \\
\hline $ R$ & 0.48 & 0.54 & 0.59 & 0.64 & 0.42 \\
 $ \%L_\text{edd}\tablenotemark{b}$ & 1.63 & 2.72 & 4.26 & 5.49 & 0.92 \\
\hline$\chi^2/\text{d.o.f.}$   & 2391.96/1703\\$\chi^2_\text{red}$   & 1.405\enddata
\tablenotetext{a}{in ph\,s$^{-1}$\,cm$^{-2}$}\tablenotetext{b}{Luminosity calculated between 0.1--300\,keV assuming a distance of 8\,kpc and a black hole mass of 10\,$\text{M}_{\odot}$}
\end{deluxetable*}

\begin{deluxetable*}{rlllll}
\tablewidth{0pc}
\tablecaption{Best-fit parameters for a free emissivity index (M1-qv).\label{tab:m1_qv}}
\tablehead{\colhead{Parameter} & \colhead{I} & \colhead{II} & \colhead{III}& \colhead{IV}& \colhead{V}}
\startdata
 $ N_\text{H}~[10^{22}\,\text{cm}^{-2}]$ & $0.794\pm0.019$ & --- & --- & --- & --- \\
 $ \text{Fe/solar}$ & $6.5\pm0.4$ & --- & --- & --- & --- \\
 $ i~[\text{deg}]$ & $43.0^{+1.5}_{-1.4}$ & --- & --- & --- & --- \\
\hline $ A_\text{cont}\tablenotemark{a}$ & $0.0611^{+0.0007}_{-0.0006}$ & $0.1008\pm0.0008$ & $0.1639\pm0.0010$ & $0.2289\pm0.0012$ & $0.0390\pm0.0004$ \\
 $ \Gamma$ & $1.545^{+0.007}_{-0.006}$ & $1.549\pm0.005$ & $1.569\pm0.004$ & $1.593\pm0.004$ & $1.572\pm0.006$ \\
 $ A_\text{refl}\tablenotemark{a}$ & $\left(1.07^{+0.35}_{-0.10}\right)\times10^{-5}$ & $\left(1.81\pm0.10\right)\times10^{-5}$ & $\left(2.63\pm0.15\right)\times10^{-5}$ & $\left(3.90\pm0.18\right)\times10^{-5}$ & $\left(0.47\pm0.04\right)\times10^{-5}$ \\
 $ \xi$ & $\left(2.04^{+0.11}_{-0.61}\right)\times10^{2}$ & $\left(2.14^{+0.08}_{-0.06}\right)\times10^{2}$ & $\left(2.35^{+0.10}_{-0.08}\right)\times10^{2}$ & $\left(2.21^{+0.06}_{-0.05}\right)\times10^{2}$ & $\left(2.12^{+0.08}_{-0.07}\right)\times10^{2}$ \\
 $ R_\text{in}~[r_g]$ & $4.3^{+0.7}_{-0.5}$ & $4.4^{+0.6}_{-0.5}$ & $4.2^{+0.5}_{-0.4}$ & $4.0\pm0.4$ & $4.0^{+0.6}_{-0.5}$ \\
 $ q$ & $5.0^{+2.0}_{-1.0}$ & $6.2^{+2.3}_{-1.2}$ & $5.4^{+1.0}_{-0.7}$ & $4.8^{+0.6}_{-0.4}$ & $1.7\pm0.5$ \\
 $ CC_\text{XRT}$ & $0.964\pm0.028$ & $0.935\pm0.015$ & $1.161\pm0.014$ & $1.138\pm0.012$ & $1.065\pm0.014$ \\
\hline $ R$ & 0.52 & 0.58 & 0.63 & 0.68 & 0.43 \\
 $ \%L_\text{edd}\tablenotemark{b}$ & 1.62 & 2.71 & 4.25 & 5.48 & 0.93 \\
\hline$\chi^2/\text{d.o.f.}$   & 2214.67/1698\\$\chi^2_\text{red}$   & 1.304\enddata
\tablenotetext{a}{in ph\,s$^{-1}$\,cm$^{-2}$}\tablenotetext{b}{Luminosity calculated between 0.1--300\,keV assuming a distance of 8\,kpc and a black hole mass of 10\,$\text{M}_{\odot}$}
\end{deluxetable*}

\begin{deluxetable}{rlllll}
\tablewidth{0pc}
\tablecaption{Best-fit parameters for the lamppost geometry (M1-LP).\label{tab:m1_lp}}
\tablehead{\colhead{Parameter} & \colhead{I} & \colhead{II} & \colhead{III}& \colhead{IV}& \colhead{V}}
\startdata
 $ N_\text{H} $ & $0.786^{+0.018}_{-0.014}$ & --- & --- & --- & --- \\
 $ \text{Fe/solar}$ & $5.27^{+0.37}_{-0.29}$ & --- & --- & --- & --- \\
 $ i [\text{deg}]$ & $40.5^{+1.1}_{-0.8}$ & --- & --- & --- & --- \\
\hline $ A_\text{cont}\tablenotemark{a}$ & $0.0609^{+0.0007}_{-0.0005}$ & $0.1002^{+0.0008}_{-0.0007}$ & $0.1626^{+0.0010}_{-0.0007}$ & $0.2275\pm0.0012$ & $0.0391\pm0.0004$ \\
 $ \Gamma$ & $1.542^{+0.007}_{-0.005}$ & $1.546^{+0.005}_{-0.004}$ & $1.565\pm0.004$ & $1.5893^{+0.0033}_{-0.0024}$ & $1.574\pm0.006$ \\
 $ A_\text{refl}\tablenotemark{a}$ & $\left(9.9544939^{+0.0000004}_{-0.8398278}\right)\times10^{-6}$ & $\left(1.71\pm0.10\right)\times10^{-5}$ & $\left(2.47^{+0.15}_{-0.14}\right)\times10^{-5}$ & $\left(3.65^{+0.17}_{-0.15}\right)\times10^{-5}$ & $\left(4.7\pm0.4\right)\times10^{-6}$ \\
 $ \xi$ & $\left(2.04^{+0.11}_{-0.53}\right)\times10^{2}$ & $\left(2.11^{+0.07}_{-0.06}\right)\times10^{2}$ & $\left(2.3243^{+0.0847}_{-0.0025}\right)\times10^{2}$ & $\left(2.2078^{+0.0556}_{-0.0021}\right)\times10^{2}$ & $\left(2.09^{+0.07}_{-0.06}\right)\times10^{2}$ \\
 $ H \text{(corona)} [r_g]$ & $2.1^{+2.1}_{-0.0}$ & $2.1^{+0.8}_{-0.0}$ & $2.1^{+0.6}_{-0.0}$ & $2.1^{+0.5}_{-0.0}$ & $2.1^{+2.7}_{-0.0}$ \\
 $ R_\text{in} [r_g]$ & $4.6^{+0.6}_{-0.5}$ & $4.3\pm0.4$ & $4.16^{+0.29}_{-0.32}$ & $4.15^{+0.25}_{-0.31}$ & $4.4^{+0.6}_{-0.5}$ \\
 $ CC_\text{XIS}$ & $0.962^{+0.028}_{-0.027}$ & $0.936\pm0.015$ & $1.162^{+0.014}_{-0.013}$ & $1.137^{+0.012}_{-0.010}$ & $1.061\pm0.014$ \\
\hline $ R$ & 0.49 & 0.55 & 0.60 & 0.65 & 0.43 \\
 $ L_x/L_\text{edd}\times100\tablenotemark{b}$ & 1.63 & 2.71 & 4.25 & 5.51 & 0.92 \\
\hline$\chi^2/\text{d.o.f.}$   & 2330.59/1698\\$\chi^2_\text{red}$   & 1.373\enddata
\tablenotetext{a}{in ph\,s$^{-1}$\,cm$^{-2}$}
\tablenotetext{b}{Luminosity calculated between 0.1--300\,keV assuming a distance of 8\,kpc and a black hole mass of 10\,$\text{M}_{\odot}$}
\end{deluxetable}

\begin{deluxetable*}{rlllll}
\tablewidth{0pc}
\tablecaption{Best-fit parameters for a free  emissivity parameter and untied photon indices (M2-qv).\label{tab:m2_qv}}
\tablehead{\colhead{Parameter} & \colhead{I} & \colhead{II} & \colhead{III}& \colhead{IV}& \colhead{V}}
\startdata
 $ N_\text{H}~[10^{22}\,\text{cm}^{-2}]$ & $0.826^{+0.020}_{-0.019}$ & --- & --- & --- & --- \\
 $ \text{Fe/solar}$ & $2.33^{+0.25}_{-0.14}$ & --- & --- & --- & --- \\
 $ i~[\text{deg}]$ & $44.3^{+2.5}_{-2.7}$ & --- & --- & --- & --- \\
\hline $ A_\text{cont}\tablenotemark{a}$ & $0.0655\pm0.0010$ & $0.1090^{+0.0013}_{-0.0012}$ & $0.1776^{+0.0018}_{-0.0019}$ & $0.2494^{+0.0022}_{-0.0027}$ & $0.0412\pm0.0005$ \\
 $ \Gamma_\text{power}$ & $1.574^{+0.009}_{-0.008}$ & $1.584\pm0.007$ & $1.605\pm0.006$ & $1.633\pm0.005$ & $1.596^{+0.007}_{-0.006}$ \\
 $ \Gamma_\text{refl}$ & $1.33\pm0.05$ & $1.337^{+0.031}_{-0.028}$ & $1.364\pm0.020$ & $1.395^{+0.014}_{-0.018}$ & $1.37\pm0.05$ \\
 $ A_\text{refl}\tablenotemark{a}$ & $\left(1.01^{+0.23}_{-0.12}\right)\times10^{-5}$ & $\left(1.73^{+0.12}_{-0.13}\right)\times10^{-5}$ & $\left(2.33^{+0.14}_{-0.19}\right)\times10^{-5}$ & $\left(3.40^{+0.17}_{-0.20}\right)\times10^{-5}$ & $\left(4.2\pm0.4\right)\times10^{-6}$ \\
 $ \xi$ & $\left(2.09^{+0.15}_{-1.04}\right)\times10^{2}$ & $\left(2.21^{+0.10}_{-0.07}\right)\times10^{2}$ & $\left(2.53^{+0.14}_{-0.10}\right)\times10^{2}$ & $\left(2.40^{+0.10}_{-0.08}\right)\times10^{2}$ & $\left(2.25^{+0.12}_{-0.09}\right)\times10^{2}$ \\
 $ R_\text{in}~[r_g]$ & $4.9^{+14.3}_{-2.8}$ & $2^{+4}_{-0}$ & $2^{+6}_{-0}$ & $2.1^{+2.5}_{-0.0}$ & $7^{+81}_{-5}$ \\
 $ CC_\text{XRT}$ & $0.949^{+0.028}_{-0.027}$ & $0.917\pm0.015$ & $1.134^{+0.014}_{-0.013}$ & $1.108^{+0.012}_{-0.010}$ & $1.052^{+0.014}_{-0.013}$ \\
\hline $ \Delta \Gamma$ & $0.25\pm0.05$ & $0.247\pm0.030$ & $0.241\pm0.021$ & $0.239\pm0.016$ & $0.23\pm0.05$ \\
 $ R$ & 0.60 & 0.68 & 0.73 & 0.79 & 0.48 \\
 $ \%L_\text{edd}\tablenotemark{b}$ & 1.63 & 2.72 & 4.30 & 5.55 & 0.93 \\
\hline$\chi^2/\text{d.o.f.}$   & 2077.77/1693\\$\chi^2_\text{red}$   & 1.227\enddata
\tablenotetext{a}{in ph\,s$^{-1}$\,cm$^{-2}$}
\tablenotetext{b}{Luminosity calculated between 0.1--300\,keV assuming a distance of 8\,kpc and a black hole mass of 10\,$\text{M}_{\odot}$}
\end{deluxetable*}

\begin{deluxetable*}{rlllll}
\tablewidth{0pc}
\tablecaption{Best-fit parameters for a free  emissivity parameter and two power-law continua (M3-qv).\label{tab:m3_qv}}
\tablehead{\colhead{Parameter} & \colhead{I} & \colhead{II} & \colhead{III}& \colhead{IV}& \colhead{V}}
\startdata
 $ N_\text{H}~[10^{22}\,\text{cm}^{-2}]$ & $0.884^{+0.012}_{-0.016}$ & --- & --- & --- & --- \\
 $ \text{Fe/solar}$ & $2.52^{+0.26}_{-0.20}$ & --- & --- & --- & --- \\
 $ i~[\text{deg}]$ & $36^{+5}_{-4}$ & --- & --- & --- & --- \\
\hline $ A_\text{cont,1}\tablenotemark{a}$ & $0.035^{+0.006}_{-0.031}$ & $0.053^{+0.014}_{-0.008}$ & $0.074^{+0.010}_{-0.014}$ & $0.091^{+0.024}_{-0.019}$ & $\le0.6\times10^{-2}$ \\
 $ \Gamma_\text{cont,1 and refl}$ & $1.433^{+0.022}_{-0.017}$ & $1.425^{+0.017}_{-0.020}$ & $1.423^{+0.010}_{-0.017}$ & $1.441^{+0.009}_{-0.010}$ & $1.391^{+0.029}_{-0.040}$ \\
 $ A_\text{refl}\tablenotemark{a}$ & $\left(0.66^{+0.10}_{-0.09}\right)\times10^{-5}$ & $\left(1.23^{+0.12}_{-0.14}\right)\times10^{-5}$ & $\left(1.64^{+0.13}_{-0.12}\right)\times10^{-5}$ & $\left(2.51^{+0.17}_{-0.16}\right)\times10^{-5}$ & $\left(3.69^{+0.30}_{-0.34}\right)\times10^{-6}$ \\
 $ \xi$ & $\left(2.38^{+0.26}_{-0.22}\right)\times10^{2}$ & $\left(2.42^{+0.14}_{-0.17}\right)\times10^{2}$ & $\left(2.95\pm0.20\right)\times10^{2}$ & $\left(2.77^{+0.23}_{-0.16}\right)\times10^{2}$ & $\left(2.50^{+0.15}_{-0.13}\right)\times10^{2}$ \\
 $ R_\text{in}~[r_g]$ & $8^{+44}_{-6}$ & $2.113^{+12.295}_{-0.004}$ & $2^{+10}_{-0}$ & $2.121^{+4.017}_{-0.012}$ & $\left(0.17^{+1.25}_{-0.15}\right)\times10^{2}$ \\
 $ q$ & $1.4^{+0.9}_{-0.7}$ & $1.71^{+0.42}_{-0.23}$ & $1.59^{+0.23}_{-0.24}$ & $1.90^{+0.19}_{-0.12}$ & $1.4^{+8.7}_{-0.9}$ \\
 $ A_\text{cont,2}\tablenotemark{a}$ & $0.035^{+0.004}_{-0.005}$ & $0.065^{+0.005}_{-0.010}$ & $0.116\pm0.008$ & $0.170^{+0.014}_{-0.010}$ & $0.0411^{+0.0006}_{-0.0024}$ \\
 $ \Gamma_\text{cont,2}$ & $1.88^{+0.07}_{-0.05}$ & $1.88^{+0.18}_{-0.05}$ & $1.858^{+0.059}_{-0.030}$ & $1.834^{+0.053}_{-0.016}$ & $1.596^{+0.013}_{-0.007}$ \\
 $ CC_\text{XRT}$ & $0.931^{+0.026}_{-0.028}$ & $0.904^{+0.017}_{-0.014}$ & $1.123^{+0.013}_{-0.015}$ & $1.102\pm0.010$ & $1.067^{+0.012}_{-0.014}$ \\
\hline $ \Delta \Gamma$ & $0.45\pm0.06$ & $0.45\pm0.12$ & $0.43\pm0.05$ & $0.39\pm0.04$ & $0.21\pm0.04$ \\
 $ R_1$ & 0.48 & 0.57 & 0.67 & 0.84 & INF \\
 $ R_{1+2}$ & 0.32 & 0.36 & 0.38 & 0.43 & 0.48 \\
 $ \%L_\text{edd}\tablenotemark{b}$ & 2.11 & 3.57 & 5.76 & 7.52 & 0.94 \\
\hline$\chi^2/\text{d.o.f.}$   & 2043.38/1688\\$\chi^2_\text{red}$   & 1.211\enddata
\tablenotetext{a}{in ph\,s$^{-1}$\,cm$^{-2}$}
\tablenotetext{b}{Luminosity calculated between 0.1--300\,keV assuming a distance of 8\,kpc and a black hole mass of 10\,$\text{M}_{\odot}$}
\end{deluxetable*}

\begin{deluxetable}{rlllll}
\tablewidth{0pc}
\tablecaption{Best-fit parameters for the lamppost geometry and two power-law continua (M3-LP).\label{tab:m3_lp}}
\tablehead{\colhead{Parameter} & \colhead{I} & \colhead{II} & \colhead{III}& \colhead{IV}& \colhead{V}}
\startdata
 $ N_\text{H}~[10^{22}\,\text{cm}^{-2}]$ & $0.908^{+0.018}_{-0.019}$ & --- & --- & --- & --- \\
 $ \text{Fe/solar}$ & $2.16^{+0.22}_{-0.18}$ & --- & --- & --- & --- \\
 $ H \text{(corona)}~[r_g]$ & $44^{+23}_{-14}$ & --- & --- & --- & --- \\
 $ i [\text{deg}]$ & $31^{+4}_{-5}$ & --- & --- & --- & --- \\
\hline $ A_\text{cont}\tablenotemark{a}$ & $0.059^{+0.004}_{-0.006}$ & $0.035^{+0.036}_{-0.020}$ & $0.120^{+0.034}_{-0.022}$ & $0.142^{+0.009}_{-0.081}$ & $0.0008^{+0.0413}_{-0.0008}$ \\
 $ \Gamma_\text{power}$ & $1.72^{+0.31}_{-0.04}$ & $1.35^{+0.06}_{-0.08}$ & $1.486^{+0.021}_{-0.142}$ & $1.496^{+0.021}_{-0.084}$ & $1.2^{+1.8}_{-0.0}$ \\
 $ \Gamma_\text{refl}$ & $1.45^{+0.08}_{-0.05}$ & $1.433^{+0.046}_{-0.024}$ & $1.415^{+0.024}_{-0.014}$ & $1.431^{+0.013}_{-0.014}$ & $1.41^{+0.04}_{-0.06}$ \\
 $ A_\text{refl}\tablenotemark{a}$ & $\left(0.55^{+0.10}_{-0.09}\right)\times10^{-5}$ & $\left(1.06^{+0.10}_{-0.16}\right)\times10^{-5}$ & $\left(1.55^{+0.15}_{-0.20}\right)\times10^{-5}$ & $\left(2.30\pm0.24\right)\times10^{-5}$ & $\left(3.4^{+0.6}_{-0.4}\right)\times10^{-6}$ \\
 $ \xi$ & $\left(2.49^{+0.27}_{-0.23}\right)\times10^{2}$ & $\left(2.51^{+0.23}_{-0.17}\right)\times10^{2}$ & $\left(3.03^{+0.25}_{-0.22}\right)\times10^{2}$ & $\left(2.87^{+0.34}_{-0.19}\right)\times10^{2}$ & $\left(2.57^{+0.18}_{-0.15}\right)\times10^{2}$ \\
 $ R_\text{in}~[r_g]$ & $\left(1.4^{+1.6}_{-1.0}\right)\times10^{2}$ & $29^{+24}_{-28}$ & $40^{+26}_{-38}$ & $2.2005^{+16.6412}_{-0.0006}$ & $\left(0.6^{+0.8}_{-0.4}\right)\times10^{2}$ \\
 $ A_\text{cont,2}\tablenotemark{a}$ & $0.011^{+0.028}_{-0.006}$ & $0.083^{+0.006}_{-0.010}$ & $0.079^{+0.020}_{-0.012}$ & $0.128^{+0.072}_{-0.028}$ & $0.0408^{+0.0009}_{-0.0059}$ \\
 $ \Gamma_\text{power,2}$ & $1.24^{+0.18}_{-0.04}$ & $1.82^{+0.09}_{-0.08}$ & $2.10^{+0.12}_{-0.14}$ & $2.00^{+0.17}_{-0.14}$ & $1.62^{+0.05}_{-0.08}$ \\
 $ CC_\text{XIS}$ & $0.933^{+0.025}_{-0.028}$ & $0.904^{+0.016}_{-0.017}$ & $1.118\pm0.018$ & $1.098\pm0.013$ & $1.066^{+0.013}_{-0.014}$ \\
\hline $ \Delta \Gamma$ & $0.49\pm0.21$ & $-0.48\pm0.11$ & $-0.61\pm0.15$ & $-0.51\pm0.17$ & $-0.4\pm1.0$ \\
 $ R$ & 0.92 & 0.68 & 0.63 & 0.72 & 1.36 \\
 $ L_x/L_\text{edd}\times100\tablenotemark{b}$ & 1.74 & 2.91 & 5.03 & 5.85 & 0.96 \\
\hline$\chi^2/\text{d.o.f.}$   & 2058.22/1686\\$\chi^2_\text{red}$   & 1.221\enddata
\tablenotetext{a}{in ph\,s$^{-1}$\,cm$^{-2}$}
\tablenotetext{b}{Luminosity calculated between 0.1--300\,keV assuming a distance of 8\,kpc and a black hole mass of 10\,$\text{M}_{\odot}$}
\end{deluxetable}

\begin{deluxetable*}{rlllll}
\tablewidth{0pc}
\tablecaption{Best-fit parameters for a free  emissivity parameter, untied photon indices and an additional narrow Gaussian line at 6.4\,keV (M2-qv-Fe).\label{tab:m2_qv_fe}}
\tablehead{\colhead{Parameter} & \colhead{I} & \colhead{II} & \colhead{III}& \colhead{IV}& \colhead{V}}
\startdata
 $ N_\text{H}~[10^{22}\,\text{cm}^{-2}]$ & $0.855^{+0.020}_{-0.016}$ & --- & --- & --- & --- \\
 $ \text{Fe/solar}$ & $1.56^{+0.10}_{-0.09}$ & --- & --- & --- & --- \\
 $ i~[\text{deg}]$ & $60^{+16}_{-9}$ & --- & --- & --- & --- \\
\hline $ A_\text{cont}\tablenotemark{a}$ & $0.0667\pm0.0010$ & $0.1116\pm0.0014$ & $0.1823^{+0.0018}_{-0.0019}$ & $0.2559^{+0.0023}_{-0.0024}$ & $0.0418\pm0.0006$ \\
 $ \Gamma_\text{cont}$ & $1.583^{+0.010}_{-0.005}$ & $1.5955^{+0.0073}_{-0.0030}$ & $1.6175^{+0.0055}_{-0.0029}$ & $1.6459^{+0.0048}_{-0.0020}$ & $1.603^{+0.008}_{-0.007}$ \\
 $ \Gamma_\text{refl}$ & $1.24\pm0.06$ & $1.27\pm0.04$ & $1.272^{+0.028}_{-0.026}$ & $1.316^{+0.019}_{-0.017}$ & $1.24\pm0.06$ \\
 $ A_\text{refl}\tablenotemark{a}$ & $\left(1.07^{+0.09}_{-0.12}\right)\times10^{-5}$ & $\left(1.76^{+0.13}_{-0.12}\right)\times10^{-5}$ & $\left(2.39^{+0.14}_{-0.16}\right)\times10^{-5}$ & $\left(3.44\pm0.18\right)\times10^{-5}$ & $\left(4.4\pm0.5\right)\times10^{-6}$ \\
 $ \xi$ & $\left(2.10^{+0.13}_{-0.10}\right)\times10^{2}$ & $\left(2.24^{+0.09}_{-0.08}\right)\times10^{2}$ & $\left(2.59^{+0.14}_{-0.12}\right)\times10^{2}$ & $\left(2.44^{+0.10}_{-0.09}\right)\times10^{2}$ & $\left(2.32^{+0.13}_{-0.10}\right)\times10^{2}$ \\
 $ R_\text{in}~[r_g]$ & $\left(0.48^{+0.57}_{-0.28}\right)\times10^{2}$ & $\left(0.6^{+1.0}_{-0.7}\right)\times10^{2}$ & $\left(0.8^{+0.4}_{-0.5}\right)\times10^{2}$ & $44^{+32}_{-21}$ & $\left(0.8\pm0.5\right)\times10^{2}$ \\
 $ q$ & $3.8^{+6.3}_{-2.1}$ & $2.4^{+7.7}_{-2.0}$ & $10^{+0}_{-8}$ & $3.2^{+6.9}_{-1.2}$ & $10^{+0}_{-8}$ \\
 $ A_{\text{FeK}\alpha}\tablenotemark{a}$ & $\left(0.82^{+0.28}_{-0.32}\right)\times10^{-4}$ & $\left(0.7\pm0.5\right)\times10^{-4}$ & $\left(1.8^{+0.4}_{-0.5}\right)\times10^{-4}$ & $\left(2.1\pm0.6\right)\times10^{-4}$ & $\left(0.54^{+0.12}_{-0.13}\right)\times10^{-4}$ \\
 $ CC_\text{XRT}$ & $0.950\pm0.028$ & $0.911\pm0.015$ & $1.124\pm0.014$ & $1.099\pm0.012$ & $1.053\pm0.014$ \\
\hline $ \Delta \Gamma$ & $0.35\pm0.06$ & $0.33\pm0.04$ & $0.345\pm0.027$ & $0.330\pm0.018$ & $0.36\pm0.06$ \\
 $ R$ & 0.66 & 0.74 & 0.81 & 0.86 & 0.53 \\
 $ \%L_\text{edd}$ & 1.64 & 2.74 & 4.37 & 5.62 & 0.94 \\
\hline$\chi^2/\text{d.o.f.}$   & 2005.72/1688\\$\chi^2_\text{red}$   & 1.188\enddata
\tablenotetext{a}{in ph\,s$^{-1}$\,cm$^{-2}$}
\tablenotetext{b}{Luminosity calculated between 0.1--300\,keV assuming a distance of 8\,kpc and a black hole mass of 10\,$\text{M}_{\odot}$}
\end{deluxetable*}

\begin{deluxetable}{rlllll}
\tablewidth{0pc}
\tablecaption{Best-fit parameters for the lamppost geometry, untied photon indices and an additional narrow Gaussian line at 6.4\,keV (M2-LP-Fe).\label{tab:m2_lp_fe}}
\tablehead{\colhead{Parameter} & \colhead{I} & \colhead{II} & \colhead{III}& \colhead{IV}& \colhead{V}}
\startdata
 $ N_\text{H} $ & $0.853^{+0.021}_{-0.013}$ & --- & --- & --- & --- \\
 $ \text{Fe/solar}$ & $1.58^{+0.10}_{-0.09}$ & --- & --- & --- & --- \\
 $ H \text{(corona)}~[r_g]$ & $2.9^{+54.8}_{-0.8}$ & --- & --- & --- & --- \\
 $ i~[\text{deg}]$ & $59^{+17}_{-9}$ & --- & --- & --- & --- \\
\hline $ A_\text{cont}\tablenotemark{a}$ & $0.0667\pm0.0011$ & $0.1114\pm0.0014$ & $0.1823\pm0.0018$ & $0.2556\pm0.0023$ & $0.0418\pm0.0006$ \\
 $ \Gamma_\text{power}$ & $1.583\pm0.009$ & $1.5944^{+0.0068}_{-0.0011}$ & $1.618^{+0.006}_{-0.004}$ & $1.645\pm0.005$ & $1.603^{+0.008}_{-0.007}$ \\
 $ \Gamma_\text{refl}$ & $1.24\pm0.06$ & $1.27\pm0.04$ & $1.279^{+0.026}_{-0.027}$ & $1.319^{+0.017}_{-0.016}$ & $1.25\pm0.06$ \\
 $ A_\text{refl}\tablenotemark{a}$ & $\left(1.08^{+1.04}_{-0.12}\right)\times10^{-5}$ & $\left(1.75^{+0.12}_{-0.11}\right)\times10^{-5}$ & $\left(2.40\pm0.16\right)\times10^{-5}$ & $\left(3.45\pm0.17\right)\times10^{-5}$ & $\left(4.4\pm0.5\right)\times10^{-6}$ \\
 $ \xi$ & $\left(2.10^{+0.13}_{-0.95}\right)\times10^{2}$ & $\left(2.24^{+0.09}_{-0.08}\right)\times10^{2}$ & $\left(2.58^{+0.14}_{-0.12}\right)\times10^{2}$ & $\left(2.44^{+0.10}_{-0.09}\right)\times10^{2}$ & $\left(2.31^{+0.12}_{-0.10}\right)\times10^{2}$ \\
 $ R_\text{in}~[r_g]$ & $39^{+32}_{-22}$ & $\left(7^{+6}_{-4}\right)\times10^{1}$ & $51^{+36}_{-26}$ & $40^{+19}_{-26}$ & $45^{+42}_{-25}$ \\
 $ A_{\text{FeK}\alpha}\tablenotemark{a}$ & $\left(7.8^{+2.3}_{-2.5}\right)\times10^{-5}$ & $\left(7\pm4\right)\times10^{-5}$ & $\left(1.7^{+0.4}_{-0.5}\right)\times10^{-4}$ & $\left(2.1\pm0.5\right)\times10^{-4}$ & $\left(5.1^{+1.2}_{-1.3}\right)\times10^{-5}$ \\
 $ CC_\text{XRT}$ & $0.949\pm0.028$ & $0.912^{+0.014}_{-0.015}$ & $1.124\pm0.014$ & $1.099\pm0.012$ & $1.052\pm0.014$ \\
\hline $ \Delta \Gamma$ & $0.34\pm0.06$ & $0.33\pm0.04$ & $0.338\pm0.027$ & $0.327\pm0.017$ & $0.36\pm0.06$ \\
 $ R$ & 0.66 & 0.73 & 0.80 & 0.85 & 0.53 \\
Eqw (FeK$\alpha$) [eV] & 21.22 & 11.59 & 17.33 & 16.17 & 23.18 \\
 $ L_x/L_\text{edd}\times100\tablenotemark{b}$ & 1.64 & 2.74 & 4.36 & 5.62 & 0.94 \\
\hline$\chi^2/\text{d.o.f.}$   & 2008.80/1692\\$\chi^2_\text{red}$   & 1.187\enddata
\tablenotetext{a}{in ph\,s$^{-1}$\,cm$^{-2}$}
\tablenotetext{b}{Luminosity calculated between 0.1--300\,keV assuming a distance of 8\,kpc and a black hole mass of 10\,$\text{M}_{\odot}$}
\end{deluxetable}

\begin{deluxetable*}{rlllll}
\tablewidth{0pc}
\tablecaption{Best-fit parameters for emissivity index $q=3$, spin $a=0$, and untied photon indices (M2-q3-a0).\label{tab:m2_q3_a0}}
\tablehead{\colhead{Parameter} & \colhead{I} & \colhead{II} & \colhead{III}& \colhead{IV}& \colhead{V}}
\startdata
 $ N_\text{H}~[10^{22}\,\text{cm}^{-2}]$ & $0.868\pm0.020$ & --- & --- & --- & --- \\
 $ \text{Fe/solar}$ & $1.73^{+0.09}_{-0.08}$ & --- & --- & --- & --- \\
 $ i~[\text{deg}]$ & $47^{+12}_{-7}$ & --- & --- & --- & --- \\
\hline $ A_\text{cont}\tablenotemark{a}$ & $0.0669^{+0.0010}_{-0.0013}$ & $0.1112\pm0.0014$ & $0.1812\pm0.0019$ & $0.2538\pm0.0023$ & $0.0420\pm0.0006$ \\
 $ \Gamma_\text{power}$ & $1.585\pm0.009$ & $1.594\pm0.007$ & $1.617\pm0.006$ & $1.643\pm0.005$ & $1.608\pm0.007$ \\
 $ \Gamma_\text{refl}$ & $1.29^{+0.07}_{-0.05}$ & $1.312^{+0.025}_{-0.033}$ & $1.333^{+0.017}_{-0.016}$ & $1.357^{+0.016}_{-0.015}$ & $1.34\pm0.04$ \\
 $ A_\text{refl}\tablenotemark{a}$ & $\left(9.6^{+1.3}_{-1.5}\right)\times10^{-6}$ & $\left(1.66^{+0.12}_{-0.11}\right)\times10^{-5}$ & $\left(2.25\pm0.14\right)\times10^{-5}$ & $\left(3.23^{+0.17}_{-0.16}\right)\times10^{-5}$ & $\left(4.2\pm0.4\right)\times10^{-6}$ \\
 $ \xi$ & $\left(2.21^{+0.22}_{-0.14}\right)\times10^{2}$ & $\left(2.28^{+0.09}_{-0.08}\right)\times10^{2}$ & $\left(2.65^{+0.14}_{-0.12}\right)\times10^{2}$ & $\left(2.52^{+0.11}_{-0.10}\right)\times10^{2}$ & $\left(2.34^{+0.12}_{-0.10}\right)\times10^{2}$ \\
 $ R_\text{in}~[r_g]$ & $\left(1.7^{+1.4}_{-1.2}\right)\times10^{2}$ & $\left(9^{+9}_{-4}\right)\times10^{1}$ & $\left(1.3^{+1.8}_{-0.6}\right)\times10^{2}$ & $65^{+56}_{-22}$ & $\left(2.3^{+0.7}_{-1.4}\right)\times10^{2}$ \\
 $ CC_\text{XRT}$ & $0.945\pm0.028$ & $0.915\pm0.015$ & $1.129\pm0.014$ & $1.105\pm0.012$ & $1.052\pm0.014$ \\
\hline $ \Delta \Gamma$ & $0.30\pm0.06$ & $0.282\pm0.029$ & $0.284\pm0.018$ & $0.285\pm0.016$ & $0.26\pm0.04$ \\
 $ R$ & 0.64 & 0.71 & 0.78 & 0.83 & 0.52 \\
 $ \%L_\text{edd}\tablenotemark{b}$ & 1.65 & 2.74 & 4.35 & 5.63 & 0.93 \\
\hline$\chi^2/\text{d.o.f.}$   & 2123.79/1698\\$\chi^2_\text{red}$   & 1.251\enddata
\tablenotetext{a}{in ph\,s$^{-1}$\,cm$^{-2}$}
\tablenotetext{b}{Luminosity calculated between 0.1--300\,keV assuming a distance of 8\,kpc and a black hole mass of 10\,$\text{M}_{\odot}$}
\end{deluxetable*}

\end{appendix}

\end{document}